\documentclass[useAMS,usenatbib]{mn2e} 
\usepackage[]{graphicx}
\usepackage[fleqn]{amsmath}
\pdfoutput=1
\usepackage{bm}
\usepackage{natbib}
\usepackage{textcomp}
\usepackage{url}
\usepackage{xcolor}
\usepackage{hyperref}
\usepackage{color}
\usepackage{soul}
\usepackage{tabularx}
\usepackage{xr}

\definecolor{refcol}{rgb}{0.3,0.,0.4}
\definecolor{refcol}{rgb}{0.,0.,0.}

\hypersetup{
  colorlinks=true,
  citecolor=refcol,
  filecolor=refcol,
  linkcolor=refcol,
  urlcolor=refcol,
  pdfborder=0 0 0.2
}

\title[GeMS review II]{Gemini multi-conjugate adaptive optics system review \nolinebreak II: \linebreak Commissioning, operation and overall performance}

\author[B.~Neichel, F.~Rigaut et al.]
{\parbox{\textwidth}{Benoit Neichel,$^{1,2}$\thanks{E-mail: \texttt{benoit.neichel@lam.fr}}
Fran\c{c}ois Rigaut,$^{3}$
Fabrice Vidal,$^{1}$
Marcos A. van Dam,$^{1,4}$
Vincent Garrel,$^{1}$
Eleazar Rodrigo Carrasco,$^{1}$
Peter Pessev,$^{1}$
Claudia Winge,$^{1}$
Maxime Boccas,$^{1}$
C\'eline d'Orgeville,$^{3}$
Gustavo Arriagada,$^{1}$
Andrew Serio,$^{1}$
Vincent Fesquet,$^{1}$
William N. Rambold,$^{1}$
Javier L\"{u}hrs,$^{1}$
Cristian Moreno,$^{1}$
Gaston Gausachs,$^{1}$
Ramon L. Galvez,$^{1}$
Vanessa Montes,$^{1}$
Tomislav B. Vucina,$^{1}$
Eduardo Marin,$^{1}$
Cristian Urrutia,$^{1}$
Ariel Lopez,$^{1}$
Sarah J. Diggs,$^{1}$
Claudio Marchant,$^{1}$
Angelic W. Ebbers,$^{1}$
Chadwick Trujillo,$^{1}$
Matthieu Bec,$^{5}$
Gelys Trancho,$^{5}$
Peter McGregor,$^{3}$
Peter J. Young,$^{3}$
Felipe Colazo,$^{6}$
Michelle L. Edwards$^{7}$}\vspace{0.4cm}\\
\parbox{\textwidth}{$^{1}$Gemini Observatory, c/o AURA, Casilla 603, La Serena, Chile\\
$^{2}$Aix Marseille Universit\'e, CNRS, LAM (Laboratoire d'Astrophysique de Marseille) UMR 7326, 13388, Marseille, France\\
$^{3}$the Australian National University, RSAA, Mount Stromlo Observatory, Cotter Road, Weston ACT 2611, Australia\\
$^{4}$Flat Wavefronts, PO BOX 1060, Christchurch 8140, New Zealand\\
$^{5}$Giant Magellan Telescope Organization Corporation, PO Box 90933, Pasadena, 
CA, 91109, USA\\
$^{6}$NASA Goddard Space Flight Center Greenbelt, MD 20771 USA\\
$^{7}$LBT Observatory, University of Arizona, 933 N. Cherry Ave, Tucson, AZ 85721, USA}}

\date{Released 2014 Xxxxx XX}

\pagerange{\pageref{firstpage2}--\pageref{lastpage2}} \pubyear{2014}

\def\LaTeX{L\kern-.36em\raise.3ex\hbox{a}\kern-.15em
    T\kern-.1667em\lower.7ex\hbox{E}\kern-.125emX}

\begin{document}

\label{firstpage2}

\maketitle

\begin{abstract}
The Gemini Multi-conjugate Adaptive Optics System - GeMS, a facility instrument mounted on the Gemini South telescope, delivers a uniform, near diffraction limited images at near infrared wavelengths (0.95 $\mu$m - 2.5 $\mu$m) over a field of view of 120$\:$\arcsec. GeMS is the first sodium layer based multi laser guide star adaptive optics system used in astronomy. It uses five laser guide stars distributed on a 60$\:$\arcsec square constellation to measure for atmospheric distortions and two deformable mirrors to compensate for it. In this paper, the second devoted to describe the GeMS project, we present the commissioning, overall performance and operational scheme of GeMS. Performance of each sub-system is derived from the commissioning results. The typical image quality, expressed in full with half maximum, Strehl ratios and variations over the field delivered by the system are then described. A discussion of the main contributor to performance limitation is carried-out. Finally, overheads and future system upgrades are described.
\end{abstract}


\section{Introduction}
\label{sec:introduction}

Adaptive Optics (AO) is a technique that aims to compensate the phase aberrations induced by atmospheric turbulence. Aberrations are measured by a Wave-Front Sensor (WFS), using observations of a Guide Star (GS). Corrections are applied by an optical active device, generally a Deformable Mirror (DM). For the current class of 8-10 meter astronomical telescopes, AO typically improves the angular resolution by an order of magnitude, and restores a resolution close to the telescope diffraction limit. Over the past 20 years, AO for astronomy has gone from a demonstration phase to a well-proven and operational technique, and it is now almost universally considered as an essential part of any new large telescope. In addition, to increase the number of targets on which AO can be used, all of the major 8 meter telescopes are now equipped with Laser Guide Stars (LGS - see e.g. \cite{wizinowich2012progress}). AO and LGS-AO observations have enabled major discoveries in astronomy with, among others, the discovery and study of the supermassive black hole at the center of our Galaxy (e.g. \cite{ghez2008gc, genzel2010gc}), detailed images of the surface of solar systems bodies (e.g. \cite{hartung2004new, depater2010persistent}), or precise morphology and dynamics of very distant galaxies (e.g. \cite{huertas2008robust, cresci2009sins, wright2009dynamics, carrasco2010}).

The advent of a new generation of AO systems called Wide Field AO (WFAO) mark the beginning of a new era. By using multiple GSs, either LGS or Natural Guide Stars (NGSs), WFAO significantly increases the field of view of the AO-corrected images, and the fraction of the sky that can benefit from such correction. Where the first AO systems (also called Single-Conjugate AO or SCAO) were well suited for observations of bright and relatively compact objects, the new generation of WFAO is opening the path for a multitude of new science studies.  

Different flavours of WFAO have been studied over the past years. They all require multiple GSs to perform a tomographic analysis of the atmospheric turbulence. What differentiates the various WFAO systems is how the turbulence correction is applied. 
Ground Layer AO (GLAO) uses a single DM optically conjugated to the telescope pupil \citep{rigaut2001glao}. If the correction is optimised over a field of view larger than the anisoplanatism angle, then only the atmospheric layers close to the ground will be compensated \citep{ragazzoni2002multiple}, providing a partial, but uniform correction over the field.  Another solution, called Multi-Conjugate AO \citep[MCAO,][]{dicke1975phase,beckers1988increasing,ellerbroek1994first,johnston1994analysis} uses several DMs optically conjugated to the main turbulence layers. In that case, all the layers close to the DM altitude conjugation will be compensated, restoring the telescope diffraction limit over field of views many times larger than the ones achievable with SCAO at Near Infra-Red (NIR) wavelengths.

MCAO for night time astronomy\footnote{MCAO systems for solar astronomy have been in use since the mid 2000s at the VTT in Tenerife and at the Dunn solar telescope at Sacramento Peak.} was first demonstrated by MAD, a prototype built at the European Southern Observatory \citep{marchetti2003mad,marchetti2007mad}. MAD used three NGSs, two DMs conjugated at the ground and at an altitude of 8.5$\:$km, and provided a corrected field of view of almost 2$\:$arcmin across. Although MAD successfully demonstrated the gain brought by WFAO over SCAO, it was limited in the number of potential targets due to limiting magnitude of the required NGSs (m$_{\rm R}<12.5$) and, essentially running out of targets, the instrument was decommissioned in 2008. The first multi-LGS WFAO system open for the community was a GLAO system operating at the MMT, which uses three 532$\:$nm Rayleigh LGSs \citep{Baranec2009glao}. GeMS, the Gemini MCAO system, is the first sodium based multi-LGS MCAO system.

This paper is the second of a review describing the GeMS project. The first paper (\cite{rigaut2013review} - hereafter Paper I) covers the first part of the history of the project, from the original idea to the first light images. It also includes a detailed description of GeMS, hence only a brief description of the system is given here. GeMS is made by two main sub-systems: (i) the LGS Facility (LGSF) that includes a 50$\:$W laser and an optical system called Beam Transfer Optics (BTO) that relays the laser light, and controls the LGSs, and (ii) the MCAO bench, called {\sc Canopus}. In short, the 50$\:$W laser is split in 5$\times$10$\:$W beams to produce the 5 LGSs projected on the sky at the corners and center of a 60\arcsec square. These LGSs feed five 16$\times$16 subapertures Shack-Hartmann WFSs (so-called LGSWFSs). The 2040 slope measurements are used to compute the MCAO high-order correction, correction provided at up to 800$\:$Hz by two deformable mirrors conjugated to 0 and 9$\:$km. In addition, up to three visible NGSs provide the measurements for the compensation of the tip-tilt and anisoplanatic modes. The tip-tilt compensation is done by a tip-tilt mirror (TTM) while the Tilt-Anisoplanatic (TA) modes are compensated by a combination of quadratic modes on DM0 and DM9. A fraction of the light from one of the NGS is directed toward a Slow Focus WFS (SFS), which controls the LGSWFS zoom to keep the instrument in focus. At the GeMS output, the corrected beam can be steered toward different science instruments attached to the Cassegrain focus instrument cluster. The main instrument used to date is GSAOI \citep{mcgregor2004gemini}, a 4k$\times$4k NIR imager covering $85''\!\times 85''$ designed to work at the diffraction limit of the 8-meter telescope. 

This paper focuses on the commissioning, overall performance and operation scheme of GeMS. The goal of this paper is to give a top-level view of the GeMS capability, that could be used for instance when preparing observations. Section~\ref{sec:commissioning} summarises the commissioning period, and details the performance of the sub-systems. Section~\ref{sec:system_verification} gives an overview of the System Verification (SV) period, and illustrates the science capability provided by GeMS. Section~\ref{sec:performance} analyses the top-level performance delivered by GeMS in term of image quality over the field and astrometry precision. Note that this paper does not intend to perform a detailed analysis of the system performance, as this will be presented in a dedicated paper. Section~\ref{sec:operations} discusses the operational scheme of GeMS, including overheads, and finally section~\ref{sec:the_future_of_gems} presents the system upgrades.

\section{Commissioning Overview} 
\label{sec:commissioning}

\subsection{Summary and timeline} 
\label{sub:summary_timeline}
In early October 2010, the decision was made to move {\sc Canopus} to the telescope and start on-sky commissioning as soon as possible. This decision was motivated by the seasonal weather conditions at Cerro Pach{\'o}n and the need for clear nights to propagate the laser for efficient commissioning. {\sc Canopus}' move from the Gemini La Serena headquarters to the telescope marked the ending of the Assembly Integration and Test (AIT) phase, and set the beginning of the on-sky commissioning period. {\sc Canopus} was installed on the telescope on January 10, 2010, and night time commissioning started on January 20, 2011. The first phase of the commissioning lasted five months, with five runs of 4 to 7 nights each. The main focus of this first period was the commissioning of LGS facility and check the {\sc Canopus} basic functionalities.

After this first commissioning period, in early June 2011, GeMS entered a planned five-month maintenance period. The Chilean winter yields conditions less favourable for AO observations, and this presented a timely opportunity to fix, repair and upgrade many GeMS systems based on the experience acquired on-sky, as well as to finish tasks that were put on hold prior to the accelerated commissioning plans starting in January 2011. 

A second period of commissioning started in November 2011, with seven runs of 5 to 9 nights spread over seven months. The objectives of this period were to demonstrate the MCAO correction, conduct GSAOI commissioning and start integrating GeMS into the Gemini science operations. In June 2012, GeMS entered its second five month shutdown phase. This engineering period was dedicated to implement the required upgrades before GeMS entered into regular operations. The latest phase of commissioning started in October 2012 with three runs of 8 nights each. 

In total, 95 nights have been used for the GeMS commissioning. Of these nights, 16 were lost to bad weather and 14 to major technical issues (defined as a problem that completely halts commissioning until it is solved). The number of major technical issues has been decreasing since December 2011, indicating that the system is getting more stable. Overall, technical issues occurred more frequently at the beginning of the runs, and were generally solved in a very short time frame by the engineering team. However this implies the need for a large engineering team either present on the summit or on-call, complicating the practical organisation of the runs. In the next sections we give more details on the commissioning of each sub-system.

\subsection{Laser Guide Star Facility commissioning} 
\label{sub:lgsf_commissioning}

The Laser Guide Star Facility (LGSF) includes the 50$\:$W laser \citep{dorgeville2002gemini,dorgeville2003laser,hankla2006twenty} and the Beam Transfer Optics \citep[BTO -][]{dorgeville2008gemini} that transports the 50$\:$W beam up the telescope, splits the beam five-ways and configures the five 10$\:$W beams for projection by the Laser Launch Telescope (LLT) located behind the Gemini South 8$\:$m telescope secondary mirror. The LGSF was the first subsystem to be commissioned. Most of the LGSF functionalities were tested and commissioned during the January to March 2011 period, however the final LGSF commissioning continued until 2012. 
An analysis of the commissioning and performance of the LGSF have been described in \cite{dorgeville2012gemini} and \cite{fesquet2013review}.

\subsubsection{Laser spot size and photon return optimisation}
\label{ssub:spotsize}

\begin{figure}
  \caption[] {An example of LGS spots constellations acquired with the full aperture Gemini acquisition camera. Natural seeing (defined at 0.5$\mu$m) is 0.65\arcsec.}
  \includegraphics[width = 1.0\linewidth]{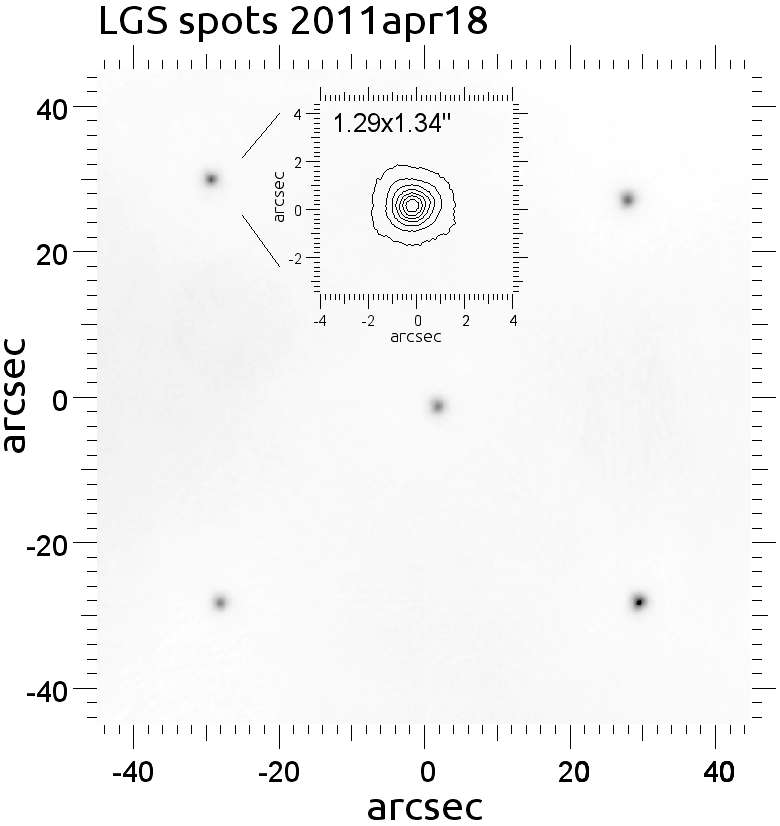}
  \label{fig:lgs_spots}
\end{figure}

Fig.~\ref{fig:lgs_spots} shows an image of the LGS constellation, acquired with the telescope Acquisition Camera (AC), when the telescope has been defocused to be conjugated to 90$\:$km. The spot Full-Width Half-Max (FWHM) is about 1.3\arcsec and almost Gaussian in shape. The natural seeing (defined at 0.5$\mu$m) was 0.65\arcsec during this acquisition. 

LGS short-exposure FWHMs obtained during the 2012 and 2013 laser runs have ranged from 1.2\arcsec (best) to 1.9\arcsec (worst) depending on seeing and actual focus optimisation, with a distribution centered near 1.7\arcsec. The original specification for the spot size was to achieve 1\arcsec FWHM on the telescope AC. The LGSWFS subaperture FoV is 2.8\arcsec, hence, spot truncation in the edge subapertures may impact the performance when seeing conditions are bad. 


The specification was built assuming a laser beam quality of M$^2$ $<$1.4, no optical aberrations induced by the BTO, and a Wave-Front Error budget of 95 nm rms for the LLT. The laser beam quality has been measured to be M$_x^2$ $\sim$1.3,M$_y^2$ $\sim$2.3, with a strong elongation due to the laser amplifiers geometry \citep{fesquet2013review}. The LLT optical quality has been tested using lucky imaging on natural stars. FWHM of 0.6\arcsec leads to a WFE of $\sim$130 nm.  We also measured that an improvement of up to 0.2\arcsec in the FWHM can be obtained when the beams are centered over the right part of the LLT, revealing issues with the LLT optical quality. Finally, on the other parameters affecting the spot size on-sky, we have measured better performance when the LLT tube covers were open, and when the air supply used in the BTO to ensure over pressure was turned off. These actions reduced the sources of turbulence and internal seeing on the BTO.

The photon return has been monitored over the period of the commissioning, from 2011 to 2013. The main variations observed are due to sodium density fluctuations as described in \cite{neichel2013sodium}. A detailed analysis of the photon return is also presented in \cite{dorgeville2012gemini}. The most important result is that the return, as measured by the LGSWFS, is a factor 2 to 5 lower than specifications. The original requirement was giving a range of 250-390$\:$ph/subaperture/frame at the design frame rate of 800Hz (the subaperture size is 50$\times$50$\:$cm$^2$), equivalent to 80 to 120$\:$ph/cm$^2$/s. Sodium return values measured during the 2011-2013 period gives number ranging from 10 to 90$\:$ph/cm$^2$/s. The lower values being observed during the low sodium season (November to February) and the highest one during the high sodium season (May to July). The impact of the low photon return on performance is discussed in Sect. \ref{sec:performance}. Several factors can explain the discrepancy between specifications and the actual results. First, the laser spectral format itself is not fulfilling the original requirements: the spectral bandwidth of the laser is twice what was designed originally. This directly and dramatically impacts the interaction with the sodium atoms, as described in e.g. \cite{moussaoui2009sodium}, \cite{holzlohner2010sodium} and \cite{rochester2012sodium}. Second, the total throughput of the BTO and LLT is 30\% below the original specification. The {\sc Canopus} throughput at 589 nm is also 50\% below the original specification. Finally, the control of the polarisation of the laser beams has not been implemented yet, which leads to relative differences of 1.5 to 2 in flux between the five beams, varying with telescope elevation.  Some of these issues will be addressed in future system upgrades as described in Sect.~\ref{sec:the_future_of_gems}.

\subsubsection{LGSF integration with CANOPUS}
\label{ssub:lgsf_vs_canopus}

During operation, once the LGSs have been acquired on the {\sc Canopus} LGSWFS, the tip-tilt error measured by each of the WFS is sent to an array of five fast steering mirrors (called FSA for Fast Steering Array) located in the BTO. This forms a closed-loop and is running at up to 200$\:$Hz. On top of compensating for possible mechanical flexure, it is also compensating for the uplink tip-tilt due to the atmospheric turbulence. It is a critical element of GeMS and the system could not work without this compensation. Closed loop performance of the fast jitter compensation by the FSA platforms gives tip-tilt residuals on the order of 0.1\arcsec Root-Mean-Square (RMS) and a closed-loop bandwidth of about 5$\:$Hz. Tip-tilt residuals are twice above the original specification, however, and because the LGS spots are bigger than originally designed, the linearity range of the LGSWFS quadcell is large enough to accommodate these residuals. For a typical spot size of 1\arcsec (measured on a subaperture), the linearity range is $\pm$0.5\arcsec. The FSA mirrors only have a small dynamical range ($\pm$5\arcsec equivalent on sky), hence their average position is offloaded to a combination of a Pointing Mirror (PM) and a Centering Mirror (CM), used to adjust both the position of the LGS constellation on-sky and the beams on the LLT. In addition, the rotation computed from the five FSA average positions is offloaded to a `K Mirror'' (KM) every 10$\:$s to compensate for constellation rotation drifts.


\subsection{CANOPUS commissioning} 
\label{sub:canopus_commissioning}

{\sc Canopus} commissioning started in March 2011, and finished in December 2011, when the first wide-field compensated images were obtained (see Paper I - Sect. 6).

\subsubsection{LGSWFS stepper look-up table} 
\label{ssub:lgs_stepper}

The LGSWFS assembly contains eight stepper mechanisms (two zoom lenses and six magnificators) used to accommodate for the changes in LGS ranges (changes in telescope elevation or changes in the Na layer altitude), as well as to compensate for flexure and {\sc Canopus} temperature variations. A LUT is built to ensure that the registration between the five LGSWFS and DM0 is maintained for all accessible LGS ranges, telescope elevation, and temperature changes. This LUT is built during day-time, using artificial laser sources located at the entrance focal plan of {\sc Canopus}. Using a model fitting method described in \citep{neichel2012identification}, we have checked that the DM9 registration was kept constant when DM0 was properly registered. 


A procedure has been developed to measure the misregistration on-sky and check the LUT performance. The method, described in \cite{rigaut2012gems}, is based on a lock-in detection of a dynamic dithering pattern introduced on DM0. Results measured on-sky show that the LUT maintain the registration below a 10\% subaperture error for the full range of elevations and temperatures seen by GeMS.

\subsubsection{Centroid gains} 
\label{ssub:centroid_gains}

As discussed in paper I (Section 4.6), centroid gains calibration have been of particular concern. This was especially critical due to the large amount of Non-Common Path Aberrations (NCPA, see Sect.~\ref{sub:ncpa}) as an error on the centroid gain will translate directly into a NCPA compensation error \citep{veran2000centroidgain}.

When using the {\sc Canopus} LGS calibration sources, because there is no turbulence, we were able to use a simple method that consists in swiping the LGS spots in front of the LGS WFS (thus through the quadcell in each subaperture) using the TT mirror. Assuming the latter is well calibrated, one obtains this way the quadcell transfer function, from which the centroid gains can be readily fitted.

On the sky however, the constraints are different. There is a large amount of natural disturbance (turbulence), and several methods have been proposed in the literature \citep{veran2000centroidgain,gratadour2007centroidgain,vandam2005centroidgain} to calibrate the centroid gains on-line. The method that has been selected for GeMS is based on dithering. A small disturbance is introduced at a given frequency. The amount of this disturbance, as detected by the WFS, is retrieved by lock-in detection. The ratio between what is introduced and what is detected is an estimator of the error on the centroid gains. 

The original plan for GeMS was to use individual rotations of each of the LGS (uplink) as a disturbance using the BTO FSA (paper I, Section 4.4.4).
This has the important advantage to have no direct impact on the image quality on the science path, thus one can possibly use fairly large dithering amplitudes, increasing the measurements SNR. Unfortunately, this was hampered by cross calibration errors between the FSA mirrors, and more importantly, by distortion of the dithering signal due to the FSA mirrors hitting their end-of-travel limits. 

A second attempt was made by introducing a similar rotating motion, but introduced by the TTM. This solved the issues encountered with the FSA method but it impacted the science image significantly. It was found that 10$\:$milli-arcsec (radius of circular motion) was the minimum quantity to get a measurement relatively immune from noise. The method was also heavily affected by narrow vibration peaks, which eventually disqualified it. 

The third disturbance type that was experimented with and eventually adopted is to induce a checkerboard-like mode using the DM \citep{rigaut2011gems}. This makes the LGS spots rotate in opposite direction in ``even'' and ``odd'' subapertures. In addition, and prior to the temporal filtering provided by the lock-in detection, a spatial projection onto this checkerboard mode is performed, providing additional filtering of noise and other natural contributors; in particular, it provides total immunity to vibration. 
A checkerboard mode amplitude of 30$\:$nm RMS is used which, at the current level of performance, does not induce any detectable effects on the images (the effect of the checkerboard mode will be to create satellites around the PSF cores). This method was cross-checked with the results obtained by dithering the tip-tilt mirror described above and the results are consistent within less than 2\%.

\subsubsection{High-order loop and related offloads} 
\label{ssub:highorderloop}
The first LGS closed loop was achieved as soon as March 2011, however, it took several more runs to be able to optimize the high-order loop and provide high-order corrections over the field. One of the issue was related to centroid gain as discussed in the previous section. Another main issue that was encountered was related to the Rayleigh contamination (or fratricide effect), as described in \cite{neichel2011sodium}. It was found that due to fast laser power variations, and spatial jitter of the beams on the LLT due to the optical location of the FSA, it was not possible to calibrate accurately enough the Rayleigh background in order to subtract it (see Paper I - Sect. 5.3.2). This leads to large LGSWFS slope errors and means that the Rayleigh-affected subapertures (about 20\% of all subapertures) have to be discarded. 

The tomographic phase reconstruction is done by the reconstructor matrix $R$. $R$ is a regularised least-squares inversion of the interaction matrix, $M$ and is given by: 
\begin{equation}
\label{eq-rec}
R = (1 - F_a)(M^T W^{-1} M + \alpha C_\phi^{-1} + \beta F_a )^{-1} M^T W^{-1}(1 - F_s)
\end{equation}
The terms in Eq.~\ref{eq-rec} are as follows:
\begin{itemize}
\item $F_a$ are the filtered modes in actuator space. For MCAO, they consist of piston, tip and tilt on the ground-layer DM, and the same modes plus focus and astigmatism for the high-altitude DM(s),
\item $W$ is a weighting matrix that weighs the centroid measurements from partially illuminated subapertures less heavily than fully illuminated ones, and also weighs smaller spots more heavily than larger ones,
\item $\alpha$ is a regularization parameter that can be configured depending on the signal-to-noise ratio,
\item $C_{\phi}$ is the covariance matrix of the actuator commands based on open-loop turbulence statistics,
\item $\beta$ is a constant adjusted to remove the filtered modes in the least-squares inversion, and
\item $F_s$ are the filtered centroids to remove average tip-tilt in each WFS. 
\end{itemize}
The interaction matrix relates the voltages on the DM actuators to the measured centroids on the WFSs. Conceptually, they are calculated by poking one actuator at a time and measuring the change in the centroids. A much faster way to measure the interaction matrix is to poke each actuator with a separate temporal frequency, take a Circular Buffer of the centroids and the actuator commands, and perform a least-squares fit of the centroids to the commands. This technique results in a fast measurement of the interaction matrix (about two minutes) and is similar to techniques used to measure the interaction matrix on-sky with system using an adaptive secondary mirror \citep{esposito2006imat}. The interaction matrix is a function of guide star altitude and is calculated every 10 km from 90 km to 140 km. For this, the interaction matrix is measured using the internal calibration sources.

The performance of the high-order loop depends on a number of parameters, but can be characterised by the residual RMS of the slopes seen by the five LGSWFSs. The residual slopes RMS include errors related to the bandwidth error (also called servo-lag), the tomography error and the noise of the five LGSWFSs. Part of the noise measured on the residual slopes is filtered by the loop, and hence should be extracted from the RMS. We estimate the noise by taking the high frequency part of the power spectra. Typical wavefront error corresponding to these LGSWFS noise corrected residual slope RMS are on the order of 350 nm, with a range spanning 200 nm to 600 nm. The original specification of the error budget was allowing less than 200 nm for the high-order terms \citep{ellerbroek2003mcao}. The main factor affecting the performance are discussed in Sect. \ref{sec:performance_limitation}.

\subsubsection{NGS loop and related offloads} 
\label{ssub:ngsloop}

The tip-tilt signal coming from the three probes (six measurements) is used to compute the weighted average tip-tilt (two modes compensated by the tip-tilt mirror), the tilt-anisoplanatic modes (three modes compensated by driving quadratic modes on the high altitude DM) and a global rotation mode (used to adjust the tracking of the Cassegrain Rotator). Thus, there are three parts to the reconstructor. 

A minimum-variance reconstructor was implemented for the tip-tilt and tilt-anisoplanatism modes based on \cite{vandam2013glao}. First, a series of points is defined where we would like to optimize the tip-tilt correction (the 'science targets'). Here, we use nine targets in a regular square grid between -30\arcsec and $+$30\arcsec. Then we estimate the tip-tilt, $\hat{s}_t$ at each point in the science field based on the tip-tilt measurements, $s_m$ using:
\begin{equation}
\hat{s}_t = C_{tm}(C_{mm} + C_{nn})^{-1}s_m,
\end{equation}
where $C_{tm}$ is the covariance matrix between the tip-tilt in the science target and the WFS directions, 
$C_{mm}$ is the covariance matrix for the tip-tilt in the WFS directions and 
$C_{nn}$ is the covariance matrix for the measurement noise.
Finally, we perform a least-squares fit to find the tip-tilt, tilt-anisoplanatism and Cassegrain rotator commands that minimize the residual errors at the science target locations.

The performance of the tip-tilt and TA loop depends on the asterism geometry and NGSs magnitude. NGS limiting magnitude is discussed on Sect. \ref{sec:performance_limitation}. At first order, the best constellations are the ones with three bright NGSs that spans the largest area of the FoV. An estimation of the tip-tilt and TA loop performance is given by the RMS of the residual NGS slopes.  A typical value for this RMS error is on the order of 15$\:$milli-arcsec, ranging from 10 to 40$\:$milli-arcsec. Original specification was giving less than 10$\:$milli-arcsec for this error term when working with bright NGS \citep{rigaut2000ttta}. This residual includes the bandwidth and the noise error, and, as in the LGS case, are compensated for the high-frequency part of the noise. Vibrations can also affect this residual error term. A very low level of vibrations has been measured using the {\sc Canopus} calibration sources, with a jitter of 6$\:$milli-arcsec RMS for tip and tilt, mainly due to peak at 55$\:$Hz induced by the cry-coolers of GSAOI \citep{rodriguez2011vibration}. On sky, more vibrations are often seen on the power spectra of the residual slopes \citep{guesalaga2012vib}. Low frequency vibrations ($<$20$\:$Hz) are believed to be due to the secondary mirror of the telescope. A large vibration peak around 85 Hz is also intermittently detected, accouting for $\sim$20$\:$milli-arcsec, the origin of this peak is unknown. The use of advanced controllers to optimally filter these vibrations is under study for GeMS \citep{guesalaga2012vib,guesalaga2013vib}. The performance of the TA loop is not affected by vibrations, however, the presence of large optical distortions in the NGSWFS focal plane prevents to close this loop during observations that would require telescope dithers larger than 10\arcsec (see Paper I - 5.3.5).

\subsubsection{Slow Focus Sensor loop} 
\label{ssub:sfsloop}

Because the LGSs are used to compensate for atmospheric focus, any changes in the sodium layer altitude cannot be disentangled from real atmospheric focus changes and will induce a focus drift. To prevent this from happening, the Slow Focus Sensor (SFS) continuously measures defocus on one of the NGSWFS. The SFS is a 2$\times$2 Shack-Hartmann WFS. The focus error measured by the SFS is used to adjust the position of the LGSWFS zoom, forming a feedback loop that compensates the focus error on the SFS, and consequently in the science focal plane (the focus flexure between the {\sc Canopus} SFS and the science instrument is essentially zero). 

The SFS loop update rate is ranging from 1 seconds to 5 minutes, depending on the GS magnitude. We have estimated that 50$\:$nm of focus (this corresponds to a loss of 4\% of Strehl ratio in H-band) corresponds to a centroiding accuracy in the SFS of 0.1 pixels. Such accuracy can be obtained with a 1$\:$s exposure time for an NGS with m$_{\rm R} < 13.0$. As a result, the current limiting magnitude for the SFS is m$_{\rm R} = 16.7$. Hence, whenever possible, the brightest of the NGS is used on the guide probe that contains the SFS, providing enough light to allow the TT/SFS split.

\begin{figure}
  \caption[] {Convergence of mean Strehl ratio over field of view (using sixteen PSFs) versus iteration number. The error bars give the RMS value of the sixteen Strehl values for each iteration.}
  \includegraphics[width = 1.0\linewidth]{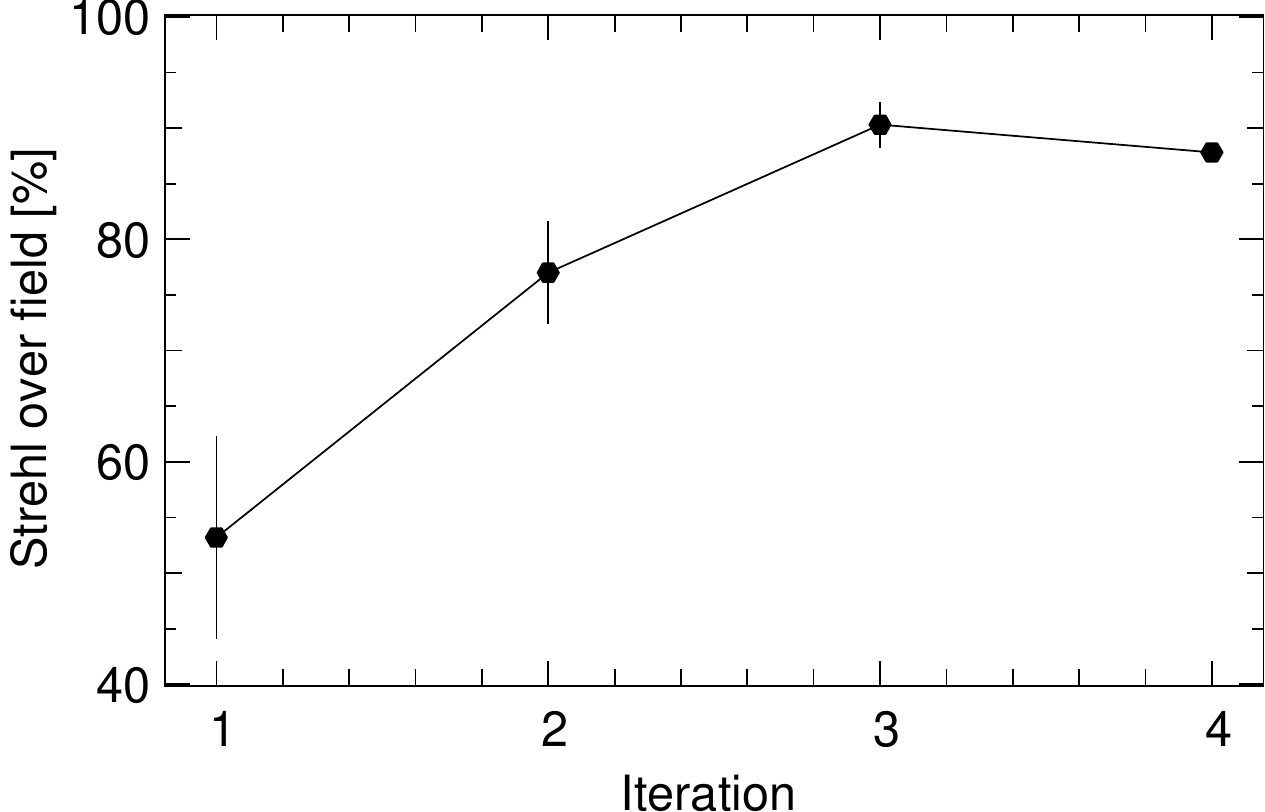}
  \label{fig:tpd_convergence}
\end{figure} 

\subsubsection{Non-Common Path Aberrations} 
\label{sub:ncpa}

The principle of NCPA compensation in GeMS has been described in details in paper I, section 4.5.2. 
Fig.~\ref{fig:tpd_convergence} shows the evolution of the field-averaged static H band Strehl ratio along the iterative NCPA optimization process. This typically uses 16 diffraction limited calibration sources spread over the GSAOI detectors. Initial Strehl ratio values vary between about 15 and 70\%. The figure shows that the process converges in 3 to 4 iterations, and reaches an average Strehl ratio of 88$\pm$0.7\%. Note that the last point shows a slightly lower average Strehl ratio that the previous iteration. However, it has a three times lower RMS, which probably explains the lower average Strehl as an important weight was given to uniformity in the minimisation criteria. The original specification was targeting for and averaged Strehl ratio of 90\%. Tests were performed when optimizing the performance in only one direction. In that case, a maximal Strehl ratio of 96\% can be reached. An important question is of course why the optimisation stops short of the single direction performance.
There are several reasons for that:
\begin{enumerate}
  \item {\bf Non-correctable aberrations}: Aberrations that are induced by optics not conjugated to one of the DM can only be partially compensated (remember the compensation has to be done over the whole field of view, not only on-axis). Only aberrations up to astigmatism are fully correctable by the DMs if they occur on optics not within the 0-9$\:$km altitude conjugation. That includes optics in the science path, but, given the procedure followed, also the optics in the WFS path (to avoid being affected by the large aberrations between the LGS WFS arms, the slope offsets at rest are absorbed in the initial slope offsets).
  \item {\bf Model calibration errors}: Any difference between the numerical model and the actual system. Misregistration error is probably the major one, but there are others.
  \item {\bf Modelling limitations}: Any effects that are not included in the numerical model and can lead to bias estimations. E.g. single wavelength image formation, non-linearities in the WFS, etc.
  \item {\bf Noise}: Science imaging detector noise, noise induced by bench local turbulence, etc.
\end{enumerate}

In the final adopted procedure, the focus is induced using DM0. This is effected using slope offsets, thus is not impacted by linearity properties of the DM. However, it will be impacted by linearity of the LGSWFS, which is another potential source of error. One image on each side of focus is generally taken, with no in-focus image (mostly for dynamical range considerations). One iteration, including two GSAOI images plus all the overhead of closing the loop, etc, takes typically 6$\:$mn.


\begin{figure*}
  \caption[]{RMC136, one of GeMS legacy images. J, H, K-band composite image. The field of view is 90\arcsec square. Averaged FWHM is 0.13$\:$\arcsec. Credit Bob Blum, NOAO.}
  \includegraphics[width = 0.8\linewidth]{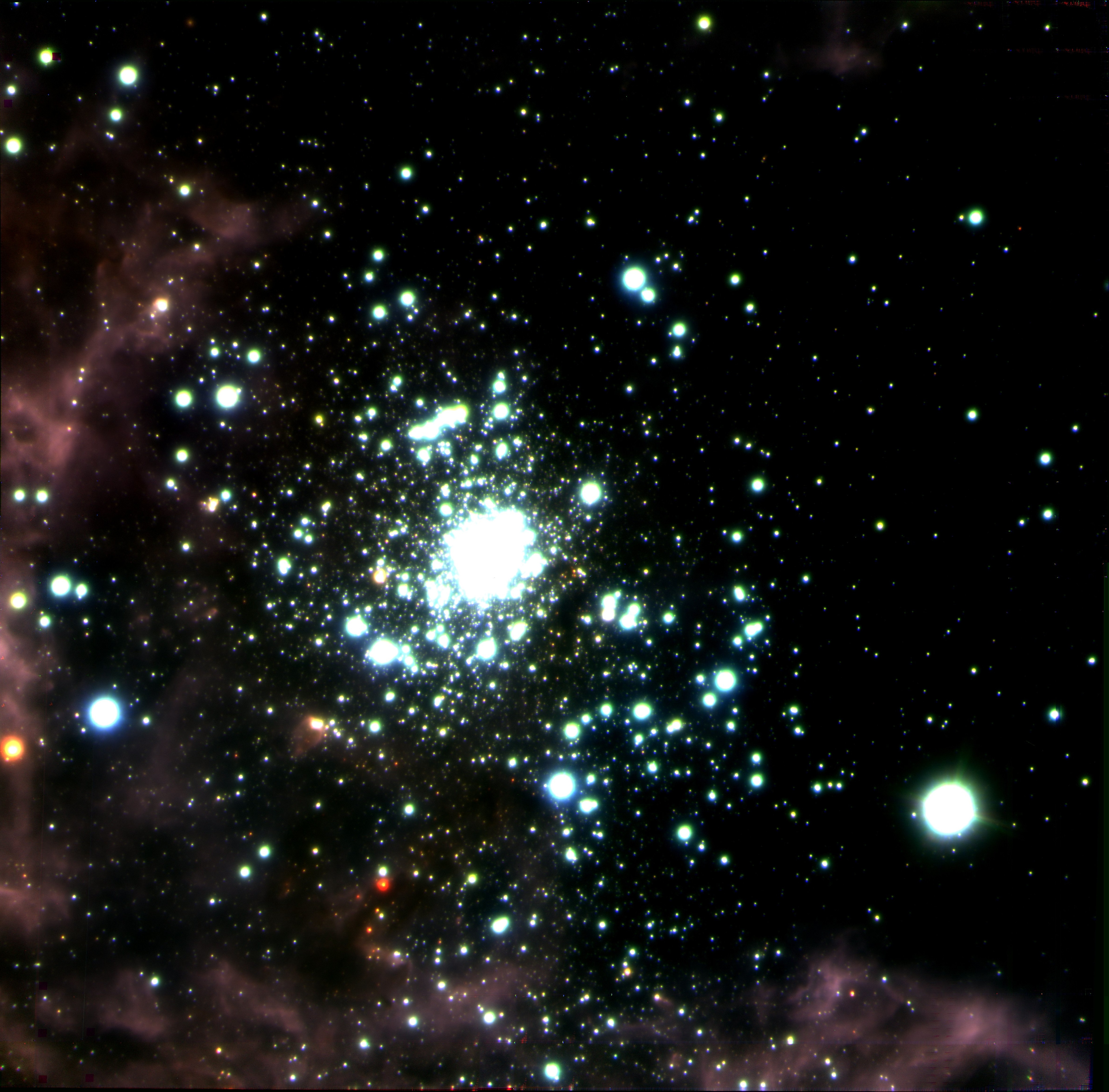}
  \label{fig:rmc136}
\end{figure*} 

\subsection{GSAOI commissioning} 
\label{sub:gsaoi}
All of the GeMS commissioning has been done using GSAOI. GSAOI is the NIR imager dedicated to GeMS \citep{mcgregor2004gemini}. It was built at the ANU. It uses four Hawaii-2RG 2k$\times$2k arrays, forming a 4k$\times$4k detector covering approximately $85''\!\times 85''$ at 0.02$\:$\arcsec per pixel. It comes with a suite of broad and narrow band filters, and has excellent image quality (H band Strehl in excess of 97\%). It is capable of Fowler sampling, and offers On-Detector Guide Window (ODGW) capabilities \citep[up to four, i.e. one per detector, see ][]{young2012odgw} to supplement or replace the {\sc Canopus} NGS TT WFS when bright enough NIR guide stars are available (see Sect. \ref{sec:the_future_of_gems}). ODGW can also be used to guide on a faint star for flexure compensation (the signal is fed to the NGS TT WFS as an offset). 

The commissioning of GSAOI per se was performed in parallel to the GeMS commissioning. A comprehensive summary of the GSAOI performance, including the characterisation of the linearity, gain and noise of the detectors, as well as the photometric zero-points, system throughput, limiting magnitude and sky brightness can be found in \cite{carrasco2012results}. For point source observations, \cite{carrasco2012results} show that a signal to noise ratio of 10 could be obtained in one hour integration for K$=$23 magnitude stars. 


\section{System Verification and shared risk period overview} 
\label{sec:system_verification}

The System Verification (SV) period started in December 2012, one year after GeMS first light, and lasted three months. The SV programs provide an end-to-end test of a new instrument or capability, from the proposal process to data delivery, prior to offering it to the community for general use. With GeMS/GSAOI, one main objective was to demonstrate the gain brought by MCAO on a large variety of science topics, including extended sources, crowded fields, and faint targets. Twenty-three programs were submitted, requesting a total of 138 hours. Of these, 13 were selected for execution between December 2012 and March 2013, for a total of 60 hours. Twelve targets out of the 13 selected were observed during the course of 18 nights\footnote{data are available at: \url{http://www.cadc-ccda.hia-iha.nrc-cnrc.gc.ca/en/gsa/sv/dataSVGSAOI_v1.html}}. The system efficiency shows that about 20\% of extra time was required to complete the programs, and that approximately 20\% of the observing time was lost due to fault. 

SV was immediately followed by a first semester of operations, offered in shared-risk mode, from March to June 2013. Around 80 hours were offered, for 11 programs, out of which 8 were completed, and 2 started (completion rate of 85\% in terms of observing hours). The system efficiency improved during this period, with only about 5\% of extra time to complete the programs, and about 10\% lost due to fault.

Fig.~\ref{fig:rmc136} shows R136, one of the target observed during the SV period. R136 is a compact star cluster located in 30 Doradus, in the Large Magellanic Cloud (LMC). Star clusters is one of the main science case for GeMS. Crowded fields are where AO/MCAO brings its largest gains. By ``compacting'' the PSF it bring out the faintest stars in the cluster which are crucial to study star formation in these environments. In addition, by delivering a uniform performance over fields that encompass most globular star cluster sizes, MCAO greatly improves the photometric precision on these crowded fields, and opens the way for a better understanding of the cluster’s stellar population, particularly of its age, any evidence for multiple stellar populations, and the distribution of low mass stars. The NGC1851 globular cluster image, presented in Fig.~\ref{fig:ngc1851} and also observed during the GeMS/GSAOI SV period, is another good illustration of the gain brought by GeMS for the star cluster science case. These data were acquired with only one NGS, located close to the center of the field, to allow for large telescope dithers (see Sect.~\ref{ssub:dither}). The average FWHM in the whole image is 95$\:$\arcsec. The effect of the tip-tilt anisokinetism can be seen on the edge of the field, however the RMS variations of the FWHM is only 12\% in this $100''\!\times100''$ image.

\begin{figure*}
  \caption[]{Abell780. Ks-filter. Field of view is 85$\:$\arcsec. Averaged FWHM is 0.077$\:$\arcsec.} 
  \includegraphics[width = 0.8\linewidth]{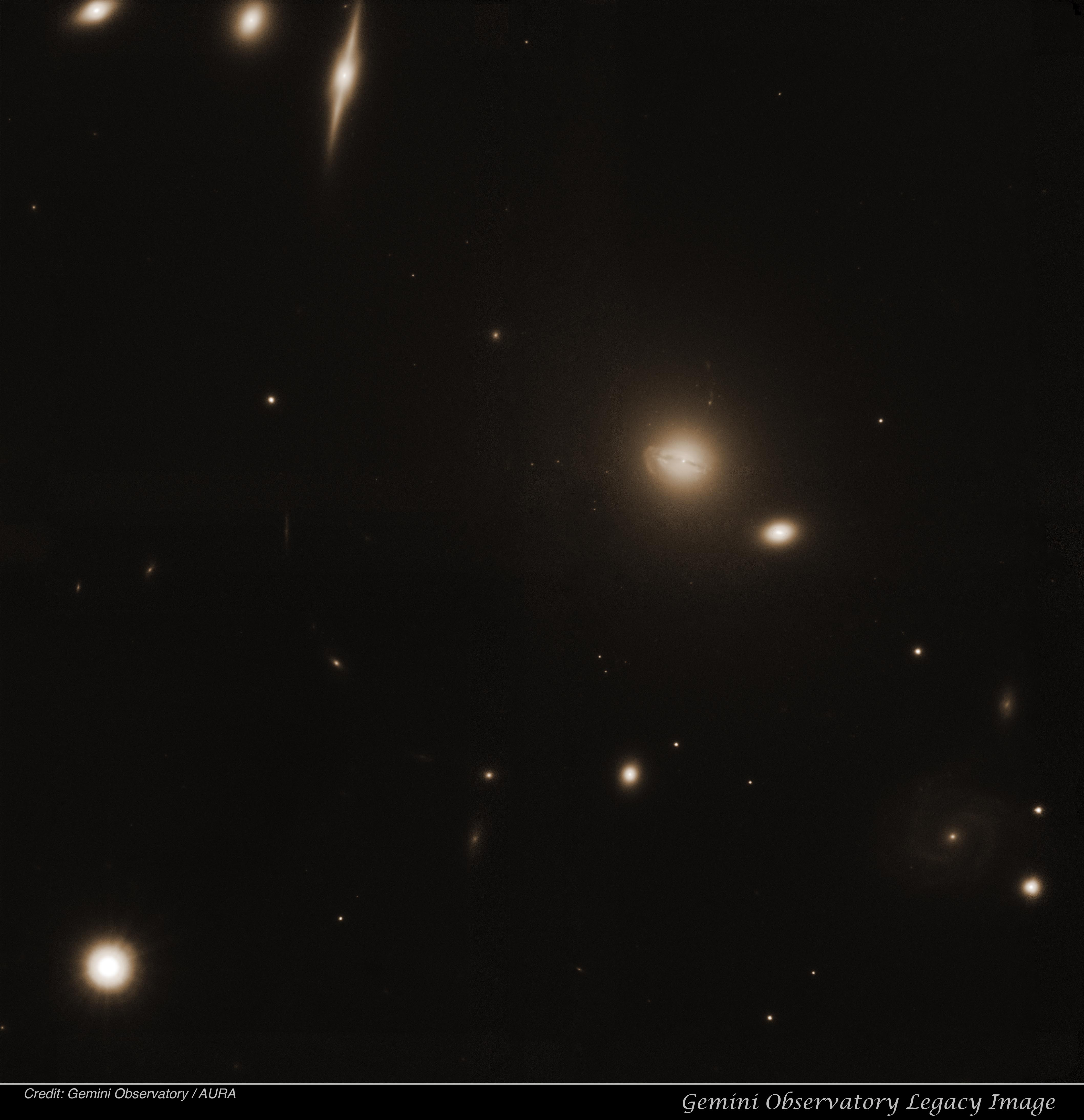}
  \label{fig:abell780}
\end{figure*}

Fig.~\ref{fig:abell780} shows another example of a target observed during the SV period; Abell 780 (better known as Hydra A) is a rich cluster of galaxies 840 million light-years distant. For this target, only two NGS have been used, one of them located on the bottom left of the image shown in Fig.~\ref{fig:abell780}, the other one located out of the field, on the top-left. Even with only two NGSs, the performance is highly uniform over the field, with an average FWHM of 0.077$\:$\arcsec. Such performance, at such distance from any usable NGS by a SCAO-LGS system, is unique to GeMS. Indeed, in a SCAO system, even when using a LGS, the target of interest must lie close enough to the NGS used for tip-tilt measurements. GeMS is using three NGSs, which can appear to be more restrictive than the SCAO-LGS mode, however, these NGSs can be anywhere in a 120$\:$\arcsec diameter acquisition field of view. Hence, the science target can be as distant as 60$\:$\arcsec from the NGS, and because of the MCAO correction, the performance will be essentially as good as if the target would have been closer to the NGS. This is particularly interesting in extra-galactic studies, which usually suffer from a low NGS and target density. 

\section{Overall performance} 
\label{sec:performance}

This section summarizes the overall performance delivered by the instrument. They are many parameters that are affecting the performance, including among others, natural seeing, NGS constellation, the number of NGSs and their brightnesses, LGS photon return (this parameter varies seasonally), turbulence speed ($\tau_0$) and profile ($C_N$$^2$), non-common path aberrations and other AO optimisation and calibration parameters. It is out of the scope of this paper to present an in-depth analysis of the performance delivered by GeMS. Instead, we focus on the averaged performance, and we identify the main contributors limiting the final results. A detailed performance analysis will be presented in a dedicated paper.

One important point regarding the performance analysis is that, for space restriction reasons in the AO bench, there are no turbulence simulators in GeMS. Hence, no formal performance characterisation has been carried out when the instruments was in the laboratory and all of the performance characterisation had to be done with on-sky data. This makes the analysis more complex as disentangling the different contributors may be difficult.The GeMS Strehl Ratios (SR) and FWHM performance presented in this section are based on data collected over 33 nights over the December 2012 to June 2013 period. An automatic tool was developed to save the data and all the environment parameters, including the AO telemetry, synchronously. This tool is described in \cite{vidal2013gems}.

\subsection{Delivered image quality} 
\label{sub:image_performance}

\begin{figure}
  \caption[]{Strehl ration (top) and FWHM (bottom) distribution versus natural seeing. K-band data are represented by the red circles, H-band by the green squares, J-band by the blue triangles. Full lines are median values of SR and FWHM for 0.1\arcsec seeing bins.}
    \begin{tabular}{c}
  \includegraphics[width = 1.0\linewidth]{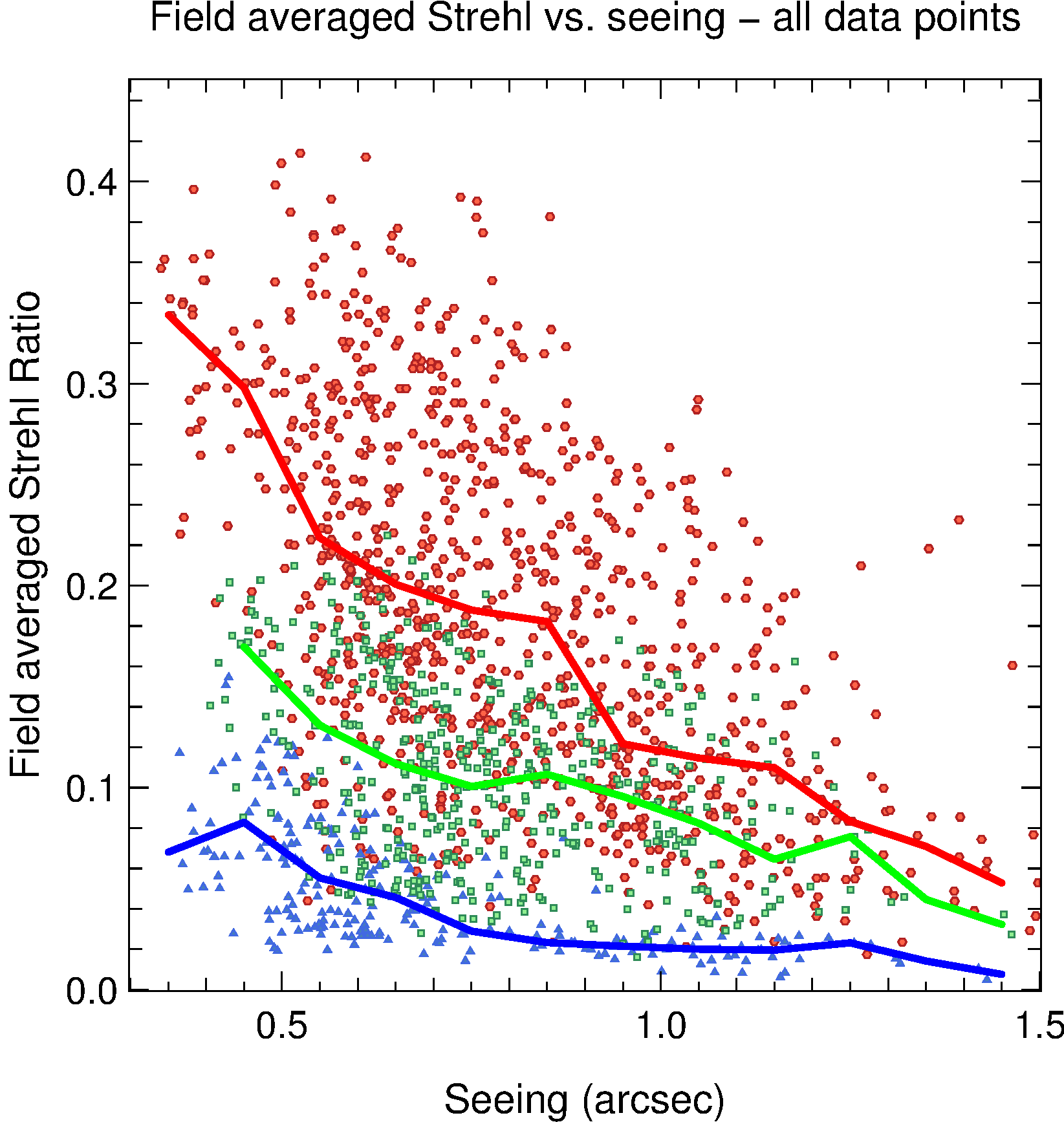} \\
  \includegraphics[width = 1.0\linewidth]{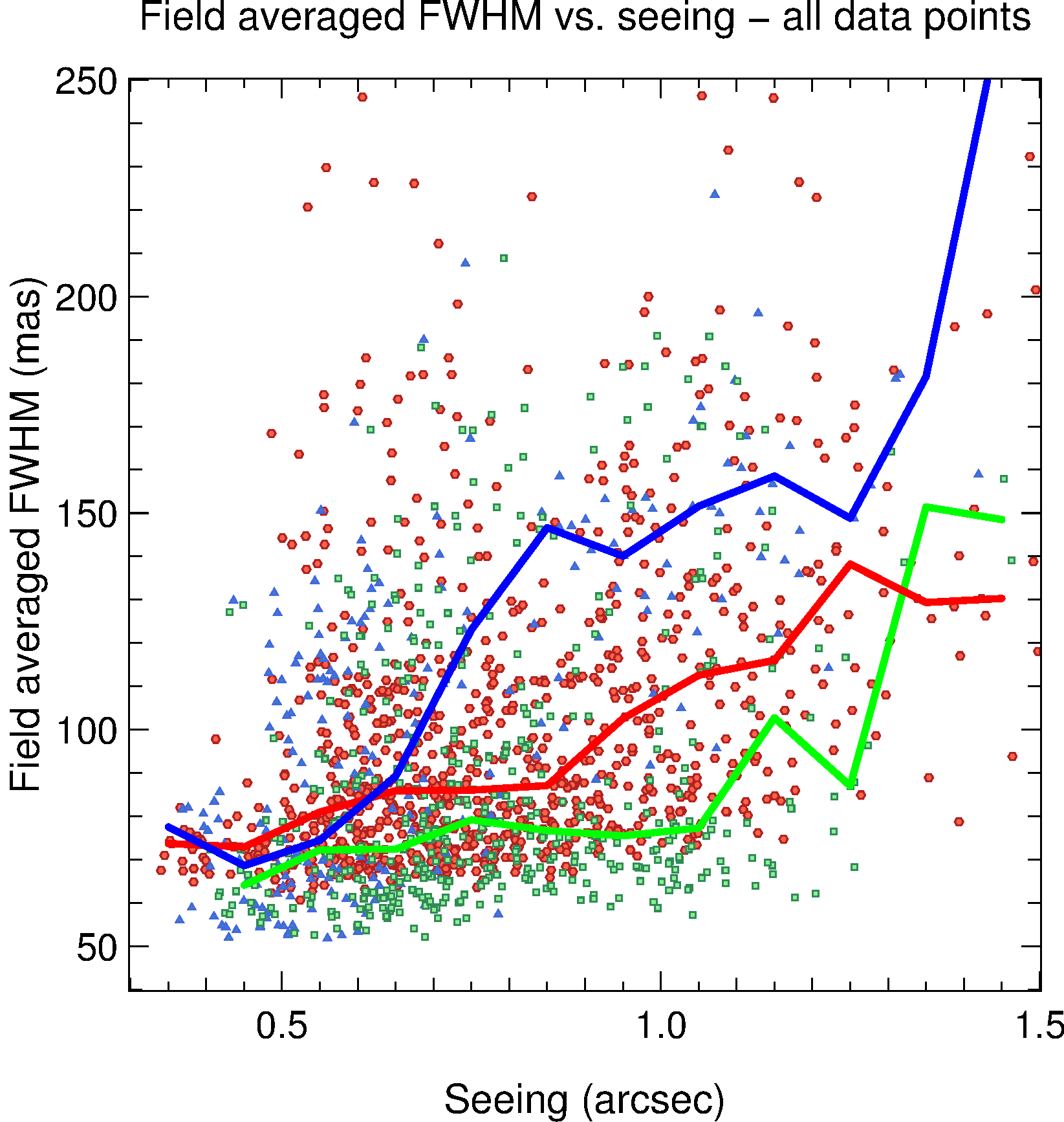}
  \end{tabular}
  \label{fig:perf1}
\end{figure}

The delivered SRs and FWHMs measured under different natural seeing conditions are shown in Fig. \ref{fig:perf1}. 
The results are based on images observed with a constellation of three NGSs and with exposure times between 10 and 180 seconds. They are respectively 950 points for the K-band images (red dots), 454 points for the H-band images (green dots) and 243 points for the J-band images (blue dots). The median natural seeing over these observations is 0.73\arcsec (defined at 0.5$\mu$m). For reference, the diffraction limited FWHM are respectively 0.031, 0.043 and 0.055$\:$\arcsec for J, H and K-band filters. 

%

Another way to present the performance is to look at the delivered image quality for a given fraction of the observing time. This is what is presented in Table~\ref{tab:gems_performance2}. Note that by observing time, we intend the time when GeMS is observing, and not the overall telescope observing time. For instance, we see that 50\% of the time, GeMS delivers a FWHM of 0.075$\:$\arcsec (or better) in H-band. In Sect. \ref{sec:performance_limitation} we discuss in more details the limitations of the current system performance.

\begin{table}
\caption{GeMS overall performance, fractional view}
\begin{center}
\begin{tabular}{l|ccc|ccc} \hline \hline
      & \multicolumn{3}{|c|}{FWHM [\arcsec]} & \multicolumn{3}{|c}{Strehl ratio [\%]} \\
Seeing conditions &    J &  H &   K &  J &  H &  K \\ \hline
20 percentile   &   0.064 & 0.064 &  0.076 &  8 & 15 & 26 \\
50 percentile   &   0.087 & 0.075 &  0.095 &  5 & 11 & 17 \\
70 percentile   &  0.110 & 0.090 & 0.110 &  3 &  8 & 13 \\ \hline \hline
\end{tabular}
\end{center}
\label{tab:gems_performance2}
\end{table}

In terms of performance uniformity over the field, MCAO brings very large gains over classical AO. Based on the data acquired during the SV, and only focusing on targets with 3 NGS, we derived an average variation of the FWHM across the images on the order of 4\% relative RMS over a field of one square arcmin. The peak to peak variations are on the order of 12\% of the average FWHM. 

\begin{figure}
  \caption[]{FWHM (top) and Strehl ratio (bottom) maps for the NGC288 target. }
  \includegraphics[width = 1.0\linewidth]{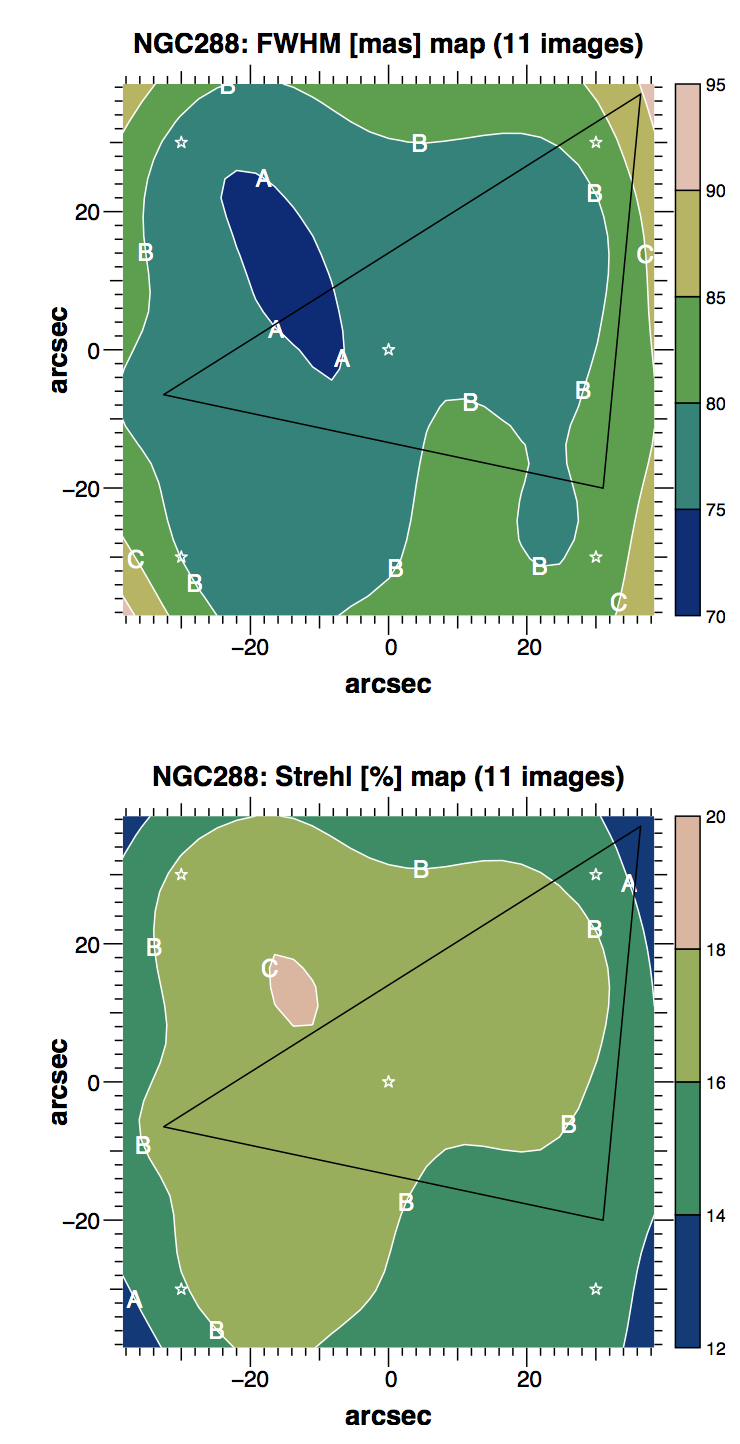}
  \label{fig:fwhm}
\end{figure}

\begin{figure*}
  \caption[]{NGC1851. Ks-filter. Field of view is 102$\:$\arcsec square. Averaged FWHM is 0.095$\:$\arcsec. From a system verification program led by Alan McConnachie. Image courtesy of Mischa Schirmer.}
  \includegraphics[width = 0.8\linewidth]{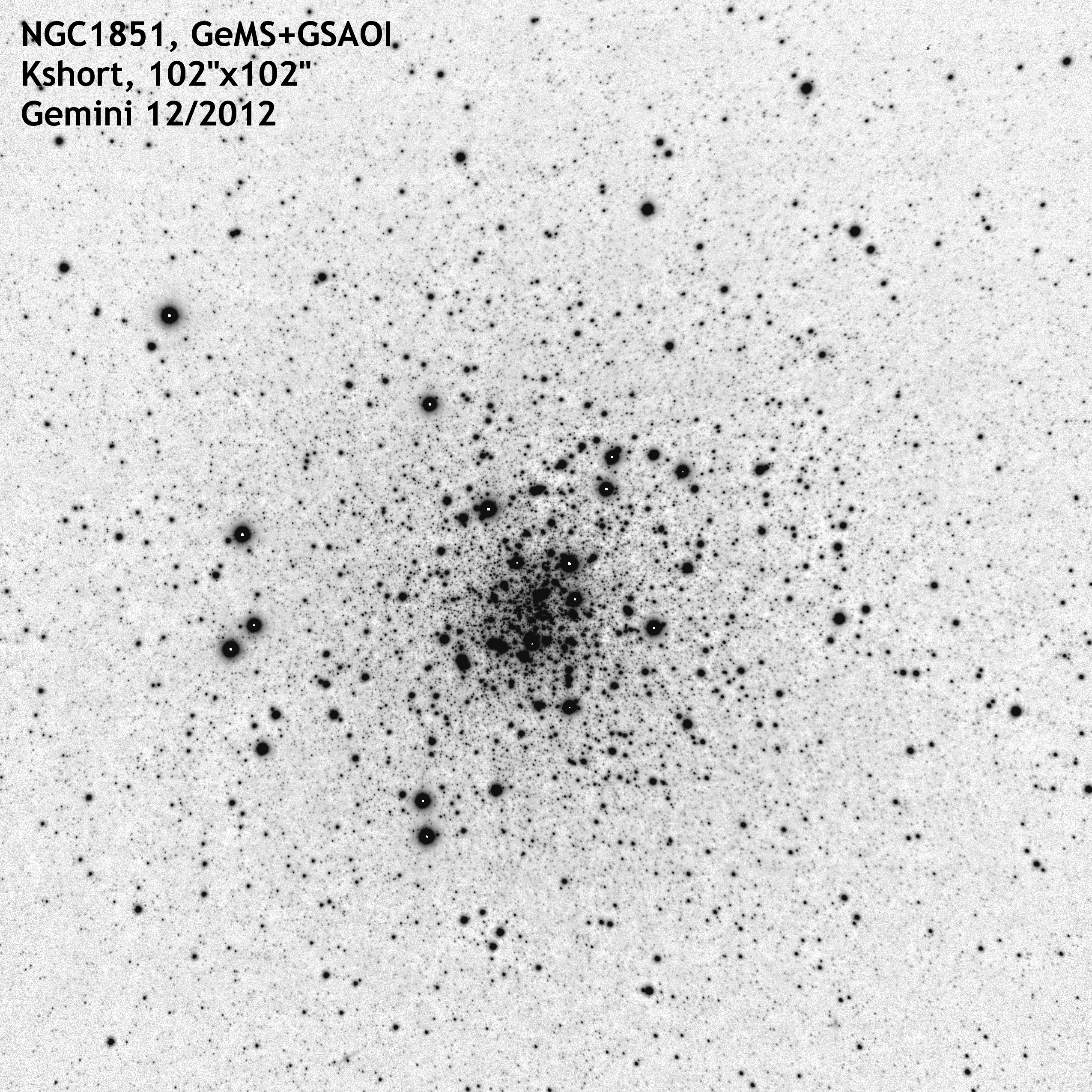}
  \label{fig:ngc1851}
\end{figure*}

Fig.~\ref{fig:fwhm} shows one example of the Strehl and FWHM distribution for the Galactic globular cluster NGC288, observed during the commissioning period (See Fig. 5 in paper I). We chose this target as performance is fairly typical. The NGS constellation used is shown with the black triangle, observations were done in the H-band filter. The average FWHM is 0.08$\:$\arcsec, the corresponding average SR is 17\%. The FWHM RMS variation over the central square arcmin is 2$\:$milli-arcsec (11$\:$milli-arcsec peak to peak). Fig.~\ref{fig:fwhm} is just one example, and a detailed analysis of the performance uniformity, and in particular its variations with the NGS constellations, will be presented in a forthcoming paper. The theoretical impact of the NGS constellation on performance variations over the field can be estimated via an algorithm presented in Sect. \ref{sub:ot_and_gems}.

\subsection{The MCAO PSF} 
\label{sub:mcaopsf}

\begin{figure}
  \caption[] {PSF profile extracted from the NGC1851 data. Top is a Lin-Log scale. Bottom is in a Lin-Lin scale.}
  \begin{tabular}{c}
  \includegraphics[width = 1.0\linewidth]{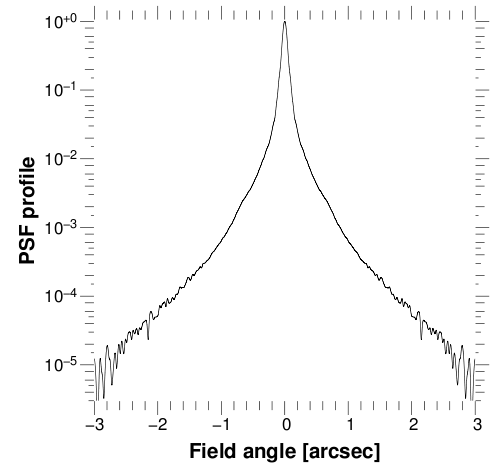} \\
    \includegraphics[width = 1.0\linewidth]{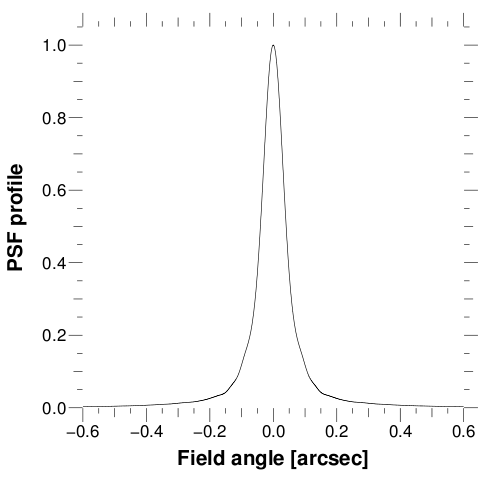}
    \end{tabular}
  \label{fig:mcaopsf}
\end{figure}

The MCAO PSF has a different shape than for regular SCAO. The main difference comes from generalised fitting \citep{rigaut2000principles}. Generalised fitting -- sometimes called generalised aliasing -- results from the fact that the phase corrector is not continuous in altitude, but instead made of a finite number of DMs, conjugated to discrete altitudes. Because the correction is effected over a finite field of view, it is impossible for the system to correct the perturbations at the DM cut-off frequency over the whole column of turbulence \citep{ragazzoni2002multiple}. As the distance of a layer to the nearest DM increases, the fitting error increases, hence the term generalised fitting. The residual phase error is thus a superposition of residual phases that have a different fitting error depending on the altitude at which they originated. This results in PSFs with Lorentzian profiles, with a narrow central core but without well defined airy rings nor the well defined dual core-halo shape usual to classical AO correction. The best functional form fitting the GeMS PSFs is:
\begin{equation} 
f(r) = g^2 / \left( g^2 + |r|^{2.4} \right)
\end{equation}
Fig.~\ref{fig:mcaopsf} shows an example of a typical PSF profile. 

\subsection{Astrometry performance}
\label{sec:astrometry_performance}

MCAO, having the ability to compensate for plate scale and dynamic atmospheric-induced field distortions, could potentially allow to reach better astrometric performance than previous AO systems. This section gives a top-level view of the astrometric performance delivered by GeMS, when using 3 bright NGS, and under typical system performance. A full analysis of the astrometric performance, including the impact of the NGS brightness and geometry, is pending.

The first astrometric performance estimation delivered by GeMS has been carried out in \cite{rigaut2012gems}. This study shows that, on a single epoch, a relative astrometric precision of 0.4$\:$milli-arcsec could be achieved for a 3 minutes total exposure in H-band. In addition, \cite{rigaut2012gems} demonstrated that the errors were coming from random sources, uncorrelated from image to image, and that no systematic error could have been detected over the course of 45 minutes. These single epoch results were confirmed by deeper observations carried-out by \citep{ammons2013astrometry} and \citep{lu2013astrometry}. \cite{lu2013astrometry}, using observations on NGC1851, demonstrated that an astrometric accuracy of 0.2$\:$milli-arcsec could be reached for stars with K $<$12 and for a total exposure time of 600 s, made by 20 exposures of 30 s. each. However, using a data set obtained over the course of 6 months, they found a systematic error of about 1$\:$milli-arcsec for multi-epoch observations. They also evidenced that due to large optical distortions in the images, best astrometric results are obtained only for non-dithered observations. 


The source of the astrometric drift for multi-epoch observations has not been identified clearly yet. The AO bench of GeMS is using a simple two Off-Axis Parabolas (OAPs) optical relay. This relay provides a clean pupil re-imaging, with little pupil distortion, but introduces a significant amount of distortions in the output focal plane. A comparison of the GeMS+GSAOI images with HST images shows a distortion pattern of up to a couple of arcseconds \citep{rigaut2012gems}. Most of the distortion pattern is static, and can be calibrated out. However, it is possible that the distortion field is evolving from one epoch to the next, impacting the ultimate astrometric performance. It might be due to changes in the gravity vector (the AO bench is mounted on the Gemini Cassegrain focus) or in the environmental parameters (temperature, humidity). As the amount of static distortion is large, even a small drift will have an impact on the final astrometric performance. In crowded fields like the Galactic Center and star clusters, the large number of stars could be enough to fit high-order polynomials to remove changing distortion. For sparse-field applications, such as using high-precision astrometry of bright stars to measure masses of orbiting exoplanets, the number of stars in the field is generally not enough to use such a method.

\subsection{Performance limitations}
\label{sec:performance_limitation}

The current GeMS performance is below the level specified in the original functional Performance Requirement Document \citep[PDR, e.g.][]{rigaut2000science}. 
Table \ref{tab:gems_performance3} summarizes the error budget described in Sect. \ref{sub:canopus_commissioning}, and compares the actual performance of the system, with the original figures from the PDR. From the PDR document, the predicted SR for an observation at 30 degree elevation, and for a median $r_0$ of 0.166 m defined at 0.55 $\mu$m, is 31\% in H-band. This translates into a global error budget of 285 nm rms. The median zenith angle measured from the actual data is 27 degree, hence close to the PDR one. Then, according to Fig. \ref{fig:perf1}, for similar seeing conditions as the PDR, the observed median H-band SR is 12\% (maximum of 20\%), which translates into a global error budget of 380 nm rms (respectively 330 nm rms). The original PDR document is distributing the error budget between three main contributors: telescope limitations, instrumental limitations and MCAO system. Each of this contributor being divided into sub-contributors. From the on-sky performance analysis, we do not have a way to disentangle the contributors of the measured error budget in order to get the same detailed analysis as in the PDR. Therefore, we have tried to combine some of the PDR contributors to match the observed error terms. We do not aim to produce a precise error budget, but rather to draw the main tendencies, and highlight the main discrepancies. The next two sections (Sect. \ref{ssub:noiseservo} and \ref{ssub:errortomo}) are giving more details on the distribution of the high-order error terms. Note that the PDR error estimation is concatenating the contribution of the low-order (NGS) and high-order (LGS) terms, but assumes a case of three bright NGS. For bright NGS, the error due to tip-tilt and tilt-anisoplanatism induces a SR reduction by a factor 0.877 \citep{rigaut2000ttta}, which translates into 95nm RMS of residual aberrations. This includes residual telescope wind-shake. The residual tip-tilt jitter measured with GeMS is 15 milli-arcsec RMS, which translates into 140 nm RMS of residual aberrations. The NCPA term estimated at PDR includes the "instrument limitations" and was evaluated to be 65nm RMS. From the results of Sect. \ref{sub:ncpa}, we evaluate the measured NCPA term to be 90nm RMS. Finally, the remaining terms of the PDR error budget are allocated to the high-order loop. This corresponds to 260nm RMS, while results from Sect. \ref{ssub:highorderloop}, gives a typical 350 nm RMS measured on-sky. All these results are summarized in Table \ref{tab:gems_performance3}.



\begin{table}
\caption{GeMS overall error budget: actual versus PDR performance.}
\begin{center}
\begin{tabular}{lcc} \hline \hline
    Error term  & Actual  & PDR \\ \hline
Total low-order (NGS) & 140 nm & 95 nm \\    
Total high-order (LGS) & 350 nm & 260 nm  \\
NCPA & 90 nm & 65 nm \\ \hline \hline
SR at H-band & 12\% & 31\% \\ \hline \hline
\end{tabular}
\end{center}
\label{tab:gems_performance3}
\end{table}

According to Table. \ref{tab:gems_performance3}, the main discrepancy between PDR and actual performance resides in the high-order terms (see Sect. \ref{ssub:highorderloop}). The high-order performance limitations can be split between noise, servo lag and tomography. The first two terms (noise and servo lag) are intrinsically linked as with GeMS, the number of photons per subaperture and per frame is usually kept constant by adjusting the high order loop frame rate. Following the method described in \cite{rigaut2012gems}, we analyse the high-order residual slopes in order to disentangle the contribution of the different error terms.

\subsubsection{Noise and servo lag error}
\label{ssub:noiseservo}
\begin{figure}
  \caption[]{Strehl ratio (top) and FWHM (bottom) distribution versus laser photon return. K-band data are represented by the red circles, H-band by the green squares, J-band by the blue triangles. Full lines are median values of SR and FWHM for photon return bins. Photon return is expressed in photons per second per meter square as measured by the LGS WFS.}
    \begin{tabular}{c}
  \includegraphics[width = 0.95\linewidth]{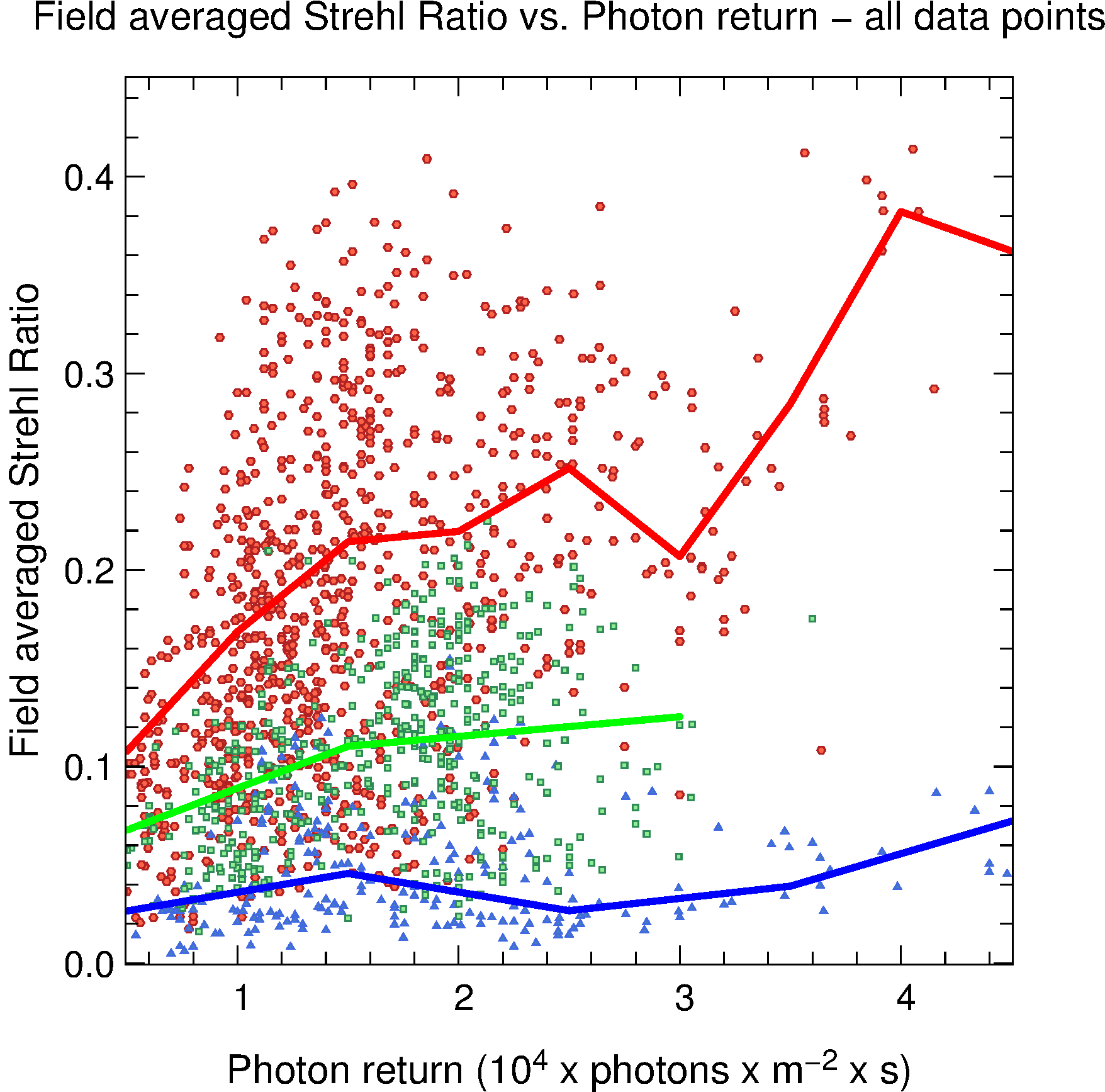} \\
  \includegraphics[width = 0.95\linewidth]{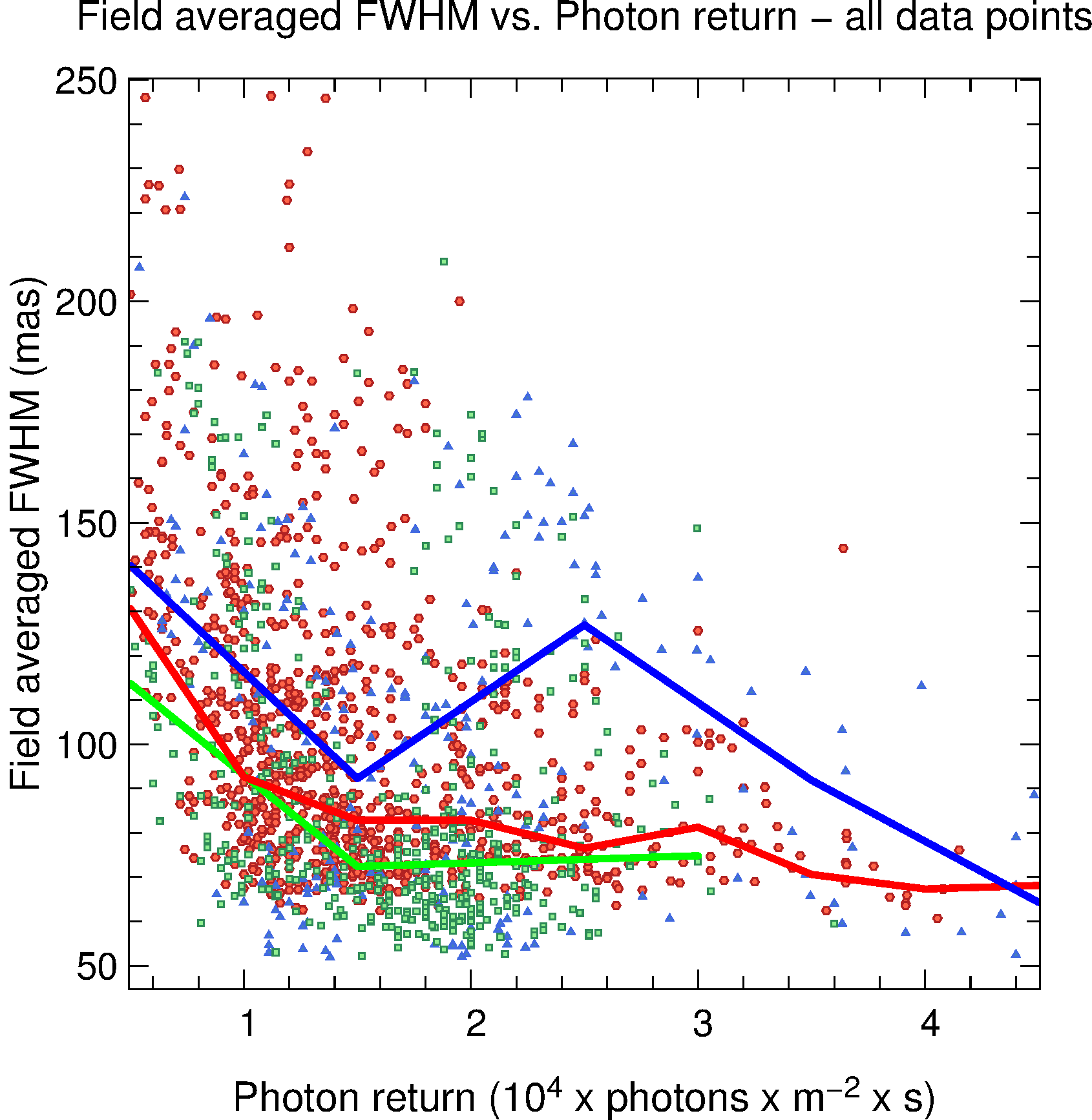}
  \end{tabular}
  \label{fig:perf2}
\end{figure}

The system is generally operated at about 140 to 160$\:$ph/subaperture/frame equivalent to between 35 and 40$\:$ph/pixel/frame, as the LGSWFSs have 2$\times$2$\:$pixels/subaperture. The LGSWFS sampling frequency is adjusted to maintain this flux level. The noise error estimated from the telemetry data is on the order of 80 nm RMS, ranging from 40 nm to 120 nm. During the low sodium season (Austral summer) a typical guide rate is about 200$\:$Hz, while during medium to high sodium season (Austral autumn to spring) the guide rate varies between 400 and 800$\:$Hz. GeMS has been design to work at a nominal frame rate of 800$\:$Hz, hence the servo lag error is often large, especially during the Austral summer. We estimated that the servo lag error was on the order of 200 nm RMS, ranging from 100 nm to 300 nm. Fig. \ref{fig:perf2} illustrates the impact of the laser photon return on the performance. As mentioned in Sect. \ref{sec:performance}, there are many parameters affecting the performance (e.g. natural seeing), which explain the large dispersion. However, and despite this dispersion, a clear tendency can be drawn, demonstrating the impact of the low photon return. 

Over the period from December 2012 to June 2013, the laser performance has been stable and the laser has delivered an average power of 42$\:$W  \citep{fesquet2013review}. This however is short of the original 50$\:$W specification, and is worsened by the lower-than-specified throughput of the BTO, LLT and Canopus, and the lower-than-specified coupling efficiency of the GeMS laser with the sodium atoms due to its spectral format \citep{neichel2013sodium}. Plans for upgrades are discussed in Sect. \ref{ssec:upgrades}.

\subsubsection{Generalised fitting (tomographic error)}
\label{ssub:errortomo}
Disentangling the impact of the tomographic error is more complex as it strongly depends on the $C_N$$^2$ profile. \cite{vidal2013gems} illustrates a case where, for a same target, same photon return and same natural seeing, the performance drops by a factor of two from one night to the next, only due to the $C_N$$^2$ distribution. Following the method described in \cite{rigaut2012gems}, we identified the tomographic error term based on the fact that the tomographic/generalized fitting error is filtered by the reconstructor and thus is not affected by the close loop transfer functions. Hence, its high frequency part (essentially, the noise) will be flat. For the data analyzed in this paper, we measure an averaged tomographic error of 280 nm RMS, ranging from 150 nm to 450 nm. Part of the tomographic error can be explained by the fact that, following technical problems with one of the deformable mirror (see paper I, Section 5.3.3), {\sc Canopus} currently uses only two DMs (at 0 and 9$\:$km with 0.5 and 1$\:$m pitch respectively) instead of the three initially planned (0, 4.5 and 9$\:$km with 0.5, 0.5 and 1$\:$m pitch respectively). This reduces the number of active actuators from 684 (design) down to 360 (current). Even though the performance will not simply scale with the number of actuators (the missing ones were aimed to deal with mid-turbulence layers where there is less turbulence than on the ground for instance), it can be easily conceived that this will reduce performance in most of the cases, and will certainly make the system performance less robust to changes in the $C_N$$^2$ profile. From simulations, we have estimated the impact on performance to be on the order of a H-band SR loss of 5\%  for typical $C_N$$^2$ profiles, but can be significantly more for unfavourable $C_N$$^2$ profiles. In effect, the current two-DM system can be viewed as a fairly potent GLAO system (17$\times$17 actuators across M1) with an additional low order DM at altitude (9$\times$9 actuator across M1).

\subsubsection{Other limiting factors}
As stated above, the main performance limitations are related to the high-order loop. On top of this, there are a number of items that are affecting performance that we try to summarize in this section.  Note that this list is not exhaustive, but includes the factors that degrade the performance the most. {\bf It is also interesting to note that these limiting factors were not foreseen during the PDR phase}. Computing the quantitative impact of each of these terms is not trivial, and we mostly focus on qualitative estimation here. 
\begin{itemize}
\item FSA limited dynamical range. When seeing is strong (e.g. $>$1\arcsec) and/or if the static optical alignment of the LGS constellation is not perfectly optimized, the FSA are often hitting their rails. This is currently the main cause of loops instabilities. This also means that the LGS spots are not properly centered, and the quadcells may be working in the non-linear regime. This effect is particularly impacting the performance as the WFS pixels are relatively small (1.38\arcsec) and the five LGS spots are not perfectly aligned with the WFS field spots. Hence vignetting with the field stop may be an issue. This latter alignment issue is intrinsic of the LGSWFS design, and could not be improved. The static constellation alignment has been improved of the years, however, due to mechanical limitations and difficult access (the optics are located behind the secondary mirror of the telescope) the best accuracy that was reached is $\sim$1.5\arcsec (equivalent on-sky) when the total dynamical range is 5\arcsec. It is difficult to quantify the impact on performance of such an effect and generally, only one out of the five LGS will go in the non-linear regime. This usually has a dramatic impact on the tomographic reconstruction, as differential aberrations between the LGSWFS will drive DM9 to the wrong shape. When this happens, strong differential elongations are seen across the science field, with amplitudes of 0.1\arcsec or more.
\item DM saturation. The original dynamical range of the DMs was $\pm$4$\mu$m. However, and due to the issues encountered with DM0, the driving voltages have been reduced by 10\% in order to try to preserve the DM life-time. This reduced the DM stroke proportionally. Moreover, protections have been implemented in order to avoid strong and recurrent DM saturation on-sky: the high-order loop is automatically opened if more than 10\% of the actuators are saturated on each DM, for more than 10 frames. In operations, and to avoid DM saturations, this implies to work with a lower loop gains and a higher loop leak than is optimal, thereby increasing the servo lag error. An approximate estimation of the impact of this error, based on the error transfer function of the system, gives around 40nm RMS.
\item Quasi-static aberrations. A large fraction of the NCPA error budget is due to aberrations introduced at the LGSWFS level. Differential aberrations at the LGSWFS level would produce a non-tomographic signal, that would be aliased in the reconstructed phase by the tomographic reconstructor. These aberrations are absorbed by the NCPA compensation, however, any drift in these aberrations, or in the on-line centroid gain estimation will lead to static or quasi-static shapes on the output science images. Such static shapes are often seen on the science PSFs, and we estimate there contributions to be $\sim$50 nm RMS. To reduce this error, an improved optical design would have included a calibration source at the LGSWFS focal plane, in order to properly calibrate these aberrations. This is however not feasible with the current LGSWFS hardware.
\item Tip-tilt loop frame rate. With the current RTC architecture, the tip-tilt loop frame rate can only be an integer of the LGS frame rate. However, and because the LGS photon return is under specification, the tip-tilt frame rate is often limited by the LGS one. In other words, there are constellations of bright NGS for which the tip-tilt loop could be run faster than the LGS one. This may explain some of the discrepancies seen between PDR and actual performance of the NGS loop reported in Table \ref{tab:gems_performance3}.
\item TA loop. As explained in Paper I - 5.3.5, the TA loop can not be closed for programs requiring telescope offsets larger than 10\arcsec. Based on simulations, we have estimated that closing the TA loop brings a gain of $\sim$3\% in H-band.
\end{itemize}


\subsubsection{NGSWFS limiting magnitude}
\begin{figure}
  \caption[]{NGS distribution for the observable sky at Gemini South: black dots are fields with three NGS, red with two, blue with one, and green with no natural guide star. The white region is the fraction of the sky not observable from Gemini South. The black dots (constellations with 3NGS) are following the galactic plane.}
  \includegraphics[width = 1.0\linewidth]{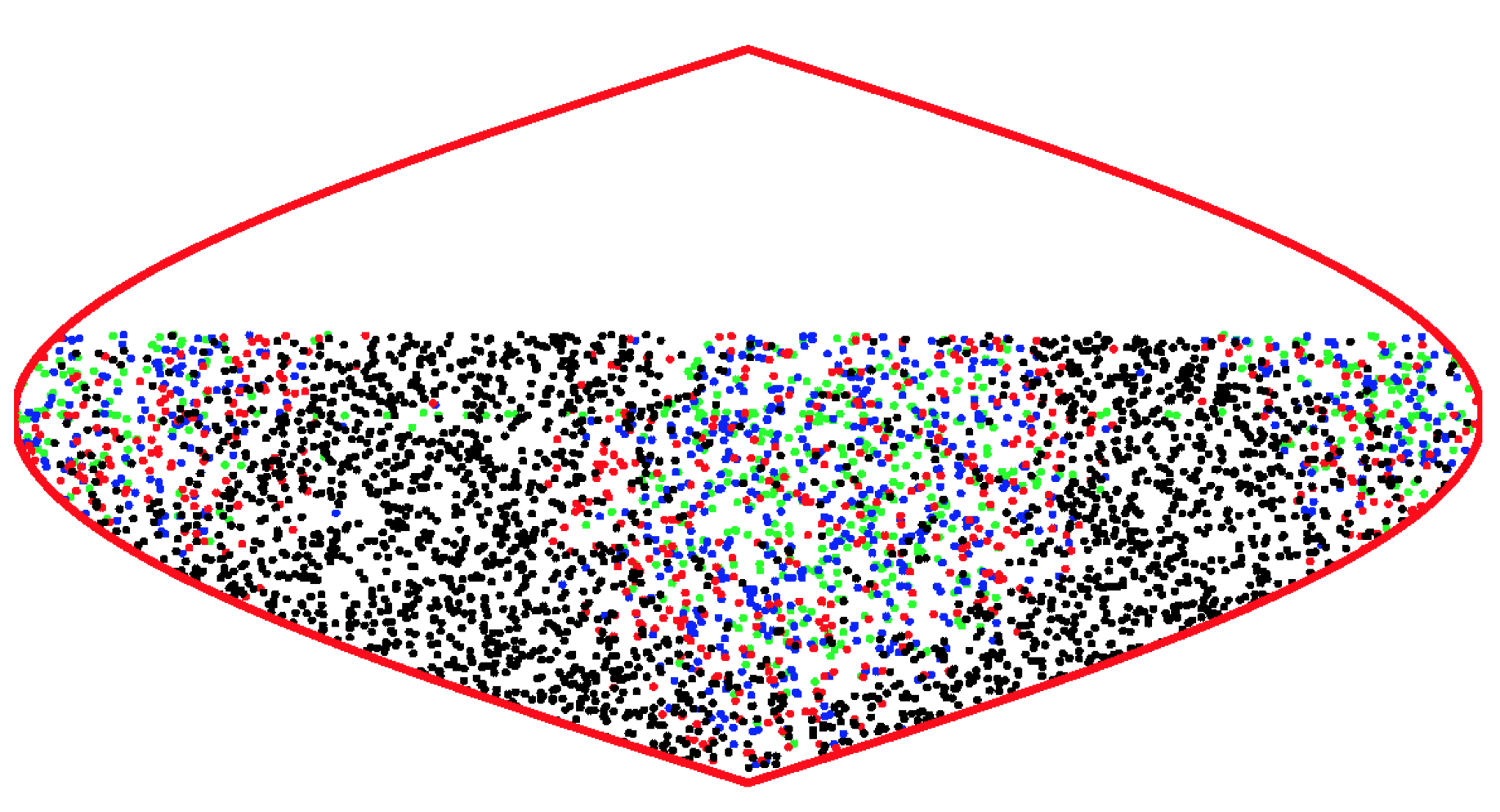}
  \label{fig:skycov}
\end{figure}

Another major issue in the current state of GeMS is the NGSWFS limiting magnitude. Due to alignment issues and design flaws, the current limiting magnitude achievable with the system is m$_{\rm R} < 15.5$. We have estimated the sky coverage achievable by GeMS by running random pointings on the portion of the sky reachable from Gemini South. Assuming a limiting magnitude of m$_{\rm R} = 15.5$, we find that the probability of having three guide stars or more is 30\%, while the probability to have no guide star at all is 35\%. Pushing the limiting magnitude to m$_{\rm R} = 18.5$ (which should be the case after the NGSWFS upgrade, see Sect.~\ref{ssub:ngs2}), 72\% of the random pointing have three NGS, and only 8\% have no guide star at all. In that case, the map in Fig.~\ref{fig:skycov} shows how the fields are distributed in the sky.

\section{Operations}
\label{sec:operations}

Despite its complexity, GeMS can be operated by a crew of only two people: the telescope operator manages all of the AO systems (on top of the telescope), while the laser operator is in charge of the laser and the BTO. The instruments are separate and are operated by the observer. 

\subsection{GeMS SMART tools \& MOP} 
\label{ssub:gems_smart_tools_&_mop}

In GeMS, there are as many as twenty loops and offloads that must be closed, monitored, and controlled. A set of dedicated tools have been developed to assist the operators in this task. The first one is called MOP, for the MYST\footnote{MYST stands for MCAO Yorick Smart Tools} Operational Panel. MOP has been designed to simplify the interaction with the AO system and integrate automation in the operation flow, assisting the operator both in the acquisition procedure and during the observation. For this, MOP gathers a limited number of possible top-level actions and status of the system in a single screen. If needed, access to low-level screens and status can be done via MOP. The SMART tools have been designed to manage the interactions between the AO system and the telescope control system. As an example, when the operator wants to close the high-order loop, a set of conditions and commands are first executed by the SMART tools: the flux is checked, low level loops are closed, default matrices are loaded.  Once the high-order loops are closed, the SMART tools continue to monitor the status and performance of all loops and can even make decisions (hopefully smart) and take actions according to the external conditions to maintain optimal performance. Optimisation of the loops is not done as a background process, but should be triggered by the operator thought MOP. This scheme was selected in order to provide more robustness to system operation. 


\subsection{Acquisition sequence and overheads} 
\label{sub:acquisition_sequence}

The multiplicity of GeMS WFSs inevitably leads to a rather complex and lengthy acquisition procedure. From the telescope slew to the beginning of the science exposure, the acquisition consists in six main steps. These steps, executed by the laser and the telescope operators, have been detailed in \cite{neichel2012science}. The overall acquisition overheads are ranging from 10 to 30 minutes, with an average of 20 minutes. For reference, the overheads associated with ALTAIR, the Gemini North SCAO system are 15 minutes in average. Most of the acquisition overheads are linked to the NGS acquisition, mainly due to the fact that the NGSWFS probes have only a 1.5$\:$\arcsec FoV. Hence for bright NGSs, in relatively sparse fields and with little catalogue errors, the full GeMS acquisition can be as fast as 10 minutes. For faint NGSs, or more complex objects (high background or a crowded field) the acquisition procedure can take over 30 minutes. The upgraded NGSWFS proposed in Section~\ref{ssub:ngs2} will greatly improve the NGSs acquisition time.

\subsection{Dither and Sky sequence}
\label{sub:dither_filter_sky}

A science observation sequence can include telescope offsets for image dithering or sky calibration, or filter changes. For all of these events, specific GeMS loops must either be paused or opened, and then resumed automatically after the event. This is handled by the SMART tools. The sequence of events for each case is described below.

\subsubsection{Dither sequence} 
\label{ssub:dither}

When a telescope offset is required, the observation sequence executor (SeqExec) sends the information to the TCS, that immediately sends it to the  SMART tools. All the NGS loops and dependencies are then paused (tip-tilt, TA, rotator and focus), while all the laser loops are kept closed. The telescope then offsets, and once all the subsystems report that they are in position, the NGS loops are resumed, and the next science exposure starts. Depending on the size of the offsets, a telescope dither can take between 3 to 30 seconds. The dither pattern is set by the NGS acquisition fields and cannot be larger than 30$\:$\arcsec.

\subsubsection{Sky sequence}

A sky sequence is somewhat similar to a dither, except that the telescope offset is usually much larger (up to 5$\:$\arcmin). In this case, all NGS loops and dependencies are paused, and the probes are frozen and do not follow telescope offsets (as a large offset would put the probes into a hard limit if they remained in follow). We found that large telescope motions could create instabilities in the high-order loop, as the LGS may be lost for a few seconds. Because of this, the high-order loop is also paused, but the LGS stabilisation loop is kept closed (FSA loop). An overhead of 60$\:$s must be accounted for.

\subsubsection{Large telescope offsets} 

If the observation requires offsets larger than 5$\:$\arcmin from the base position, the laser propagation must be stopped due to Laser Clearing House (LCH) restrictions (see Sect. \ref{ssub:lch}). When the telescope returns to the original base position, the laser operator must re-acquire the LGS. Moreover, since this offset will also be unguided, the telescope operator must re-check the NGS acquisition and correct for any telescope pointing errors. In this case a separate observation block is required and extra overheads are introduced.

\subsection{Observation interruptions}
\label{sub:observation_interruption}

\subsubsection{LCH Predictive Avoidance} 
\label{ssub:lch}

To prevent any laser illumination of sensitive satellite optics, all laser targets must first be pre-approved by the United-State space command Laser Clearing House. A week before a laser run, Gemini provides the coordinates of each target and receives in return a file from the LCH showing the allowed observation windows for this target. To maximise time on-sky and minimise any potential for inadvertent illumination, an automatic software that handles the laser shutters based on the LCH data has been developed. Only the LGS-related loops are affected. NGS loop and its dependencies do not need to be paused. The current overheads associated for a LCH window are on the order of 30$\:$s to 1 mn and also depends on the phasing of the window with the observation sequence. 

\subsubsection{Aircraft Avoidance} 

In the case of an aircraft passing within 25 degrees of the laser propagation area, the same procedure of pause/resume as the LCH windows is used, except that in this instance the laser operator must manually pause the laser propagation. While an isolated aircraft event will have a larger overhead due to the time required for an aircraft to pass through the safety zone, overall aircraft avoidance overheads are only a factor of 1.3 times larger than satellite overheads as there are fewer aircrafts passing near the propagation zone than satellites.

\subsection{Elevation and weather limitations} 
\label{sub:operational_limitations}

They are several limitations imposed by GeMS in terms of operations. The main ones are listed below.
\subsubsection{Limitation in elevation}
The elevation range accessible by GeMS goes from 40 to 85$\:$degrees. This  limitation is imposed by the LGSWFS zoom mechanism, which cannot mechanically keep the LGS in focus for elevations higher than 85$\:$degrees or lower than 40$\:$degrees. In addition, the LGS return flux decreases significantly at high airmass which also affects the performance. Objects that transit near to zenith also impose a limitation. The speed at which the laser constellation can rotate is limited by the rotation speed limit of the BTO K-mirror. For instance, for an object transiting at a peak elevation of 80 degrees, the ``dead zone'' spans $\pm$2 minutes about transit.
\subsubsection{Clouds and high-wind}
Because of safety concerns, the laser cannot be propagated in clouds that could hide planes from the laser spotters. The laser can be propagated in thin cirrus, however these cirrus will reduce the laser photon return and affect the AO performance. High winds buffeting the telescope will create wind shake and can jeopardise the tip-tilt loop stability. We found that the tip-tilt loop could survive with wind on the secondary of up to 2.5 - 3.0$\:$m/s.


\subsection{Integration with the observatory tools} 
\label{sub:ot_and_gems}

The Gemini Observing Tool (OT) is the software used to define the instrument configuration and pre-plan the observations from approved proposals. 

To provide the users with a quantitative selection criteria for the required NGSs, an automatic algorithm called {\sc Mascot} has been implemented in the OT \citep{trancho2008gemini}. This algorithm finds the best asterism\footnote{\textit{a prominent pattern or group of stars that is smaller than a constellation}, Oxford dictionary} of three stars in the group of N stars located inside the NGS patrol field of a specific science field \citep{flicker2002tilt,ellerbroek2001methods}. We define here best asterism as the asterism that will provide the highest and most homogeneous correction level over the science field of view. 
The algorithm can also handle sub-optimal cases, for example, it can return the best two-star asterism if a three-star asterism is not available. Finally, the user can also overwrite the {\sc Mascot} output, and select manually the guide stars to use for a given observation.

\begin{figure}
  \caption[]{The OT position editor. The GSAOI detectors (cyan four large squares) and the {\sc Canopus} patrol field area (red circle) are marked. The three CWFSs and the ODGW stars are represented by green squares. The Strehl map from the best asterism is superimposed as green/yellow contours. The average, RMS, minimum and maximum Strehl ratio and FWHM values are shown at the bottom of the position editor.}
  \includegraphics[width = 1.0\linewidth]{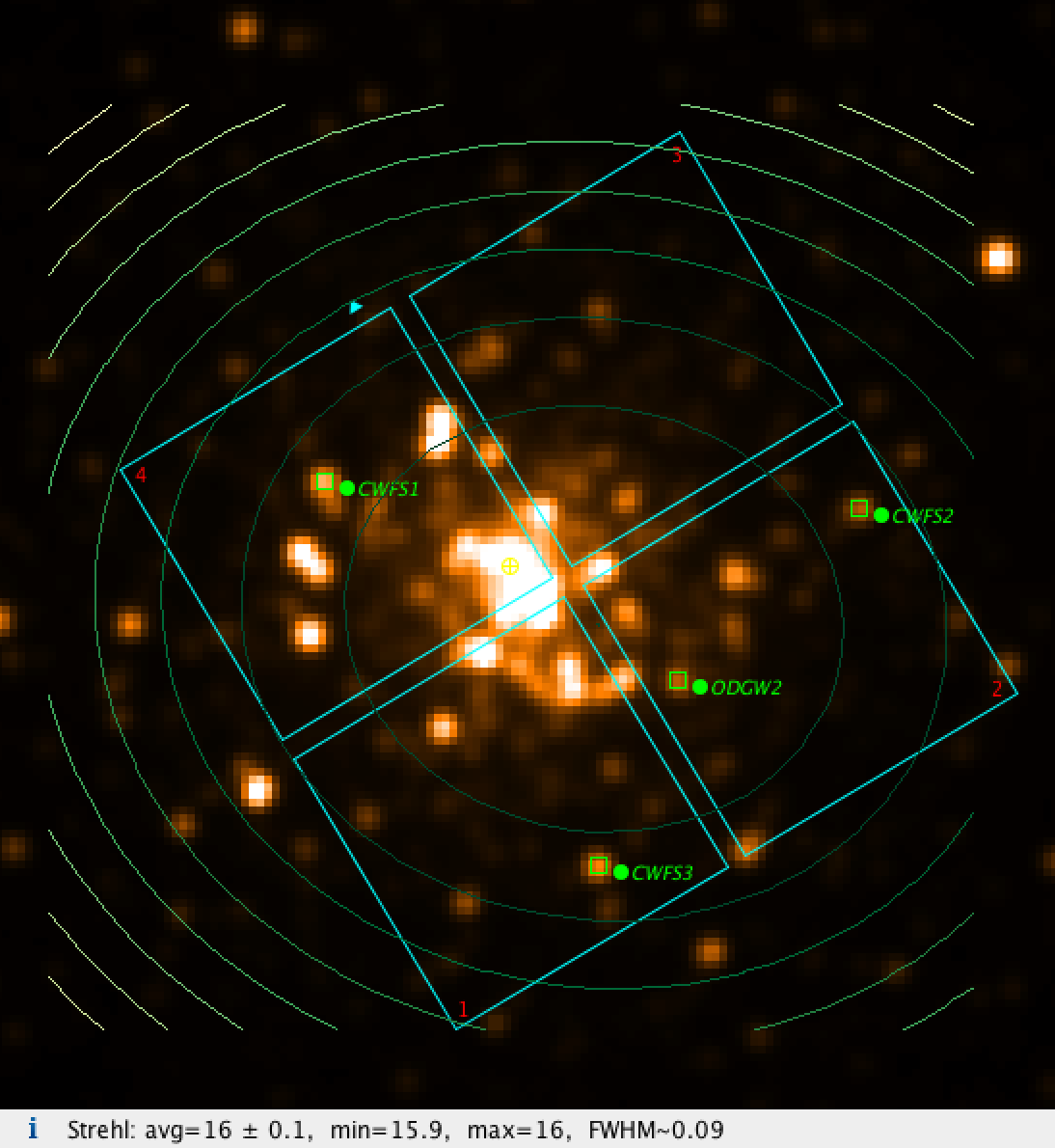}
  \label{fig:ot3}
\end{figure}

The selected asterism can be visualised in the OT position editor. Fig.~\ref{fig:ot3} shows an example for the NGC1851 field. The three selected NGS stars are marked as a small green square, and labelled as {\sc Canopus} WFS (CWFS) 1 to 3. The flexure star, which in that case is a GSAOI ODGW star (see Sect. \ref{ssub:flexloop}), is also marked with a green square, and labelled ODGW2, as the star falls in quadrant 2 of GSAOI. The average SR and FWHM, as well as the variation over the field, calculated using the {\sc Mascot} algorithm, are shown at the bottom of the position editor. Predicted iso-Strehl contours are shown as narrow curved green/yellow lines. 


\section{The future of GeMS} 
\label{sec:the_future_of_gems}

It is expected that GeMS will be in regular operation, as a facility instrument, in November 2013. The first mode of operation offered is to combine GeMS with GSAOI. More instruments should be commissioned with GeMS in the following semesters. In addition, there are a number of items that have not been commissioned yet, and will possibly be in the near future. Finally, it is planned to improve GeMS performance and address the issues discussed in Sect.~\ref{sec:performance_limitation}. These upgrades will be implemented as they become available.

\subsection{New science capabilities}

\subsubsection{GeMS + FlamingosII}

\mbox{Flamingos-2} \citep[GMOS,][]{elston2003performance} is a NIR imager, long-slit and Multi-Object Spectrograph (MOS). The main scientific interest of GeMS+F2 being the MOS mode, as GSAOI already provides the imaging capability. F2 has been designed to work both in seeing limited, and diffraction limited mode. When used in conjunction with GeMS, it has a 0.09$\:$\arcsec pixel scale, and covers a 2$\:$\arcmin FoV. The commissioning of GeMS+F2 should start in the course of 2014.

\subsubsection{GeMS + GMOS}
During the commissioning of GeMS, we had the opportunity to obtain data with the Gemini Multi-Object Spectrograph \citep[F2,][]{hook2004gmos} in March and May 2012. GMOS is a spectro-imager working in the visible. In its spectrograph mode, long-slit, multi-slit and one IFU are available. 
The imaging FoV of GMOS through GeMS covers an un-vignetted field of 2.5 arcmin$^2$, with a pixel scale of 35.9 milli-arcsec.
The three CCD chips form a 6144 x 4608 pixel array, with two gaps of about 37 pixels separating the detectors. With GeMS, only the reddest GMOS filters can be used. These are the i-band (706-850 nm), CaT-band (780-933 nm) and z-band ($\geq$ 848 nm). A first analysis of the performance delivered by GeMS/GMOS is presented in \cite{hibon2013gmos}. Although it was not originally intended to offer the GeMS/GMOS combination, it is now planned to open this mode in the future semesters.

\subsection{Missing commissioning items}

\subsubsection{Atmospheric Dispersion Compensators (ADCs)}
{\sc Canopus} has two ADCs. One is located on the science path, and can be inserted or removed from the path, the other one is in the NGSWFS path, and is always in. None of these ADCs have been commissioned yet. The current strategy for the NGS ADC will be to set it at the beginning of an observation based on the average telescope elevation, and then keep it fixed to avoid pupil wandering in the SFS path. In the worst case of an observation spanning from Zenith to 40$\:$degree of elevation, the maximum relative elongation due to atmospheric dispersion is $\pm 0.3$$\:$\arcsec 
for optical wavelength (550 to 850$\:$nm). The impact on the SNR is then small compared to the complexity of having this ADC in follow mode. The science ADC will be commissioned along with the coupling of GeMS with Flamingos-2.
\subsubsection{Flexure loop}
\label{ssub:flexloop}
The flexure loop uses the signal coming from an On-Instrument (OI) WFS to compensate for potential differential flexure between the AO bench and the instrument. The flexure signal is used as soon as the tip-tilt loop is closed by moving the three tip-tilt WFS probes all together. As the tip-tilt loop is closed on the signal coming from these probes, moving the probes moves the TTM accordingly, and results in image motion at the science detector level. First results obtained on sky have shown that the differential flexures between the AO bench and (for instance) GSAOI are small. We measured image motion on the order of $0.2''$ on sky for 30 degree elevation steps. Taking into account that the longest exposure time used with GSAOI are on the order of five minutes (to avoid saturation by sky background), we computed that the maximal SR loss due to flexure during a single exposure could never be greater than 1\% at H-band. Hence, the flexure correction, although tested during commissioning, has never been implemented in the operational scheme. The gain provided was too small compared to the complexity it entailed. When using with Flamingos-2, and because the exposure times will increase, this assessment might change. 
\subsubsection{ODGW fast guiding}
When using GeMS with GSAOI, the ODGWs can be used in a fast read out mode, providing the tip-tilt information based on the centroid position of the stars. This mode is particularly interesting for targets embedded in dust, where NIR stars are more easily achievable than the visible ones.  In this mode, one visible star (the {\sc Canopus} probe C3) is still needed for slow focus compensation. This mode has been successfully tested during the commissioning, however has not yet been integrated into operation. It will be offered if enough scientific programs require it. 


\subsection{System upgrades}
\label{ssec:upgrades}

\subsubsection{NGSWFS}
\label{ssub:ngs2}
 As explained in paper I - Sect. 5.3.4, 
  the current NGS WFS has a very low sensitivity. A dedicated attempt to fix the design has been unsuccessful. A project is currently under way at the Australian National University (ANU) to build a replacement for the TT NGS WFS, based on a single focal plane array covering the whole 2 arcmin field of view, reading out at 400Hz with less than 2 electrons read out noise. This will boost significantly the TTWFS performance; nominally back to the expected level (limiting magnitude of 18.5). It will also drastically ease the acquisition procedure, and the need for lengthy probe mapping calibrations. Finally, it should provide a solution to the distortion issue discussed in paper I (Sect. 5.3.5). The new NGSWFS is currently build at ANU, and should be commissioned at Gemini towers the end of 2014.
  \subsubsection{DM4.5}
  The replacement for DM0 should arrive in the first half of 2014, and should be re-integrated in {\sc Canopus} during the same year. 
  \subsubsection{Laser}
  The 589$\:$nm LMCT laser is relatively stable at about 42 to 45$\:$W, which is adequate for the high sodium season but significantly limits performance at other times. It is also quite demanding in term of maintenance. Although there is no definite plan, there has been discussions of a possible upgrade. This might be especially timely considering new technologies like Raman lasers \citep{feng200925} or optically pumped semiconductor lasers \citep{berger2012towards}.
  \subsubsection{RTC}
  The RTC has been generally quite reliable. However, it is an ageing piece of equipment (about 10 years old) and as noted in Paper I, can not accommodate a Pseudo Open-Loop control model which would certainly improve both performance and stability. An upgrade of the RTC is thus desirable. However, such an upgrade would entail both new hardware and new software, and will inevitably be a major and expensive endeavour. There is no current official plan to replace the RTC.
  \subsubsection{Laser beam quality}
  The 50$\:$W laser beam shape, depending on the alignment, can vary from run to run. It is often out of specification in term of beam quality (M$^2$), which result in larger or aberrated laser spots on the sky, and reduces the SNR on the LGS WFS measurements. Andr\'es Guesalaga from the Pontificia Universidad Cat\'olica de Chile and his team are currently building a two-deformable mirror system to remedy that situation. This system should allow to improve significant the laser beam quality and is to be inserted in the BTO \citep{bechet2013two}. It should be tested in 2015.
  \subsubsection{Astrometry calibration}
  As discussed in Sect.~\ref{sec:astrometry_performance}, the current astrometric performance is limited by the large amount of quasi-static distortions present in the science path. 
A possibility to improve the astrometric calibration could be to add a diffractive grid in the optical path, which will generate a grid of diffracted ``stars'' from the primary target star (see \cite{guyon2012astrometry,bendek2012astrometry,ammons2013astrometry}).  These stars will be numerous enough to fit and remove changing optical distortion. A proposal has been submitted to develop such an hardware solution for GeMS. Funding is currently under investigation.


\section{Conclusion}
\label{sec:conclusion2}
GeMS has proven to be an exciting and challenging project. Because many complex subsystems must work in unison, GeMS imposes a level of complexity rarely found in an observatory facility instrument. The amount of time on sky (about 100 nights) for commissioning activity and work towards the transition into regular operations is testament alone to the complexity of the system. Operating such a complex instrument is also a challenge for the Gemini observatory and a dedicated effort had been put in place in order to simplify and stream-line operations. 

GeMS is also a pathfinder instrument, paving the way for the future developments of such systems. It is important to emphasize that no show stoppers have been found on the way, although the delivered performance differs from what was predicted during the design phases. The compensation performance is currently limited by servo lag, noise and generalised fitting, but other, and unexpected terms are also affecting the performance. Furthermore, several calibrations designed during the early phase of the project proved to be either not working or sub-optimal during the commissioning, and new solutions had to be developed. As discussed in this paper, work is continuing on the system to improve on all these aspects, as well as many others.

But if GeMS is complex, is it also a rewarding instrument. The programs executed to date and associated images have already demonstrated its huge and unique scientific potential. Images with FWHM of 0.08$\:$\arcsec or better over the 85$\:$\arcsec$\!\times$ 85$\:$\arcsec field of view are typically obtained under median seeing or better. 

\section*{Acknowledgments}

GeMS was a large instrumentation project. In the course of the last 13 years, it involved people from many different disciplines. The authors would like to recognise the contribution and thank 
Claudio Arraya, 
Corinne Boyer, 
Fabian Collao, 
Paul Collins, 
Felipe Daruich, 
Herman Diaz, 
Matt Doolan, 
Brent L. Ellerbroek, 
Aurea Garcia-Rissmann, 
Fred Gillett, 
Alejandro Gutierrez, 
Mark Hunten, 
Stacy Kang, 
Matteo Lombini, 
Bryan Miller, 
Matt Mountain, 
Rolando Rogers, 
Roberto Rojas, 
Jean-Ren{\'e} Roy, 
Michael Sheehan, 
Doug Simons, 
Jan van Harmelen, 
and Shane Walker.

The {\sc Mascot} algorithm has been developed in collaboration with Damien Gratadour from Observatoire de Paris. 

Part of this work has been funded by the French ANR program WASABI - ANR-13-PDOC-0006-01.

Based on observations obtained at the Gemini Observatory, which is operated by the 
Association of Universities for Research in Astronomy, Inc., under a cooperative agreement with the NSF on behalf of the Gemini partnership: the National Science Foundation (United States), the National Research Council (Canada), CONICYT (Chile), the Australian Research Council (Australia), Minist\'{e}rio da Ci\^{e}ncia, Tecnologia e Inova\c{c}\~{a}o (Brazil) and Ministerio de Ciencia, Tecnolog\'{i}a e Innovaci\'{o}n Productiva (Argentina).

In memory of Vincent Fesquet, GeMS' optical and laser engineer and our friend.


\bibliographystyle{mn2e}

\bibliography{ao,gems}

\begin{thebibliography}{}

\bibitem[\protect\citeauthoryear{Ammons, Bendek, Guyon, Marois, Neichel,
  Galicher \& Macintosh}{Ammons et~al.}{2013}]{ammons2013astrometry}
Ammons M.,  Bendek E.,  Guyon O.,  Marois O.,  Neichel B.,  Galicher R.,
  Macintosh B.,  2013, in AO4ELT3 On-sky pathfinder tests of calibrated mcao
  astrometry and implications for mcao on elts

\bibitem[\protect\citeauthoryear{Baranec, Hart, Milton, Stalcup, Powell,
  Snyder, Vaitheeswaran, McCarthy et~al.,}{Baranec
  et~al.}{2009}]{Baranec2009glao}
Baranec C.,  Hart M.,  Milton N.,  Stalcup T.,  Powell K.,  Snyder M.,
  Vaitheeswaran V.,  McCarthy D.,    et~al., 2009, The Astrophysical Journal,
  693, 1814

\bibitem[\protect\citeauthoryear{B{\'e}chet, Guesalaga, Neichel \&
  Guzman}{B{\'e}chet et~al.}{2013}]{bechet2013two}
B{\'e}chet C.,  Guesalaga A.,  Neichel B.,    Guzman D.,  2013, in AO4ELT3 A
  2-deformable-mirror concept and algorithm to improve the laser efficiency of
  {G}emini {S}outh {MCAO}

\bibitem[\protect\citeauthoryear{Beckers}{Beckers}{1988}]{beckers1988increasing}
Beckers J.~M.,  1988, in Hulrich M.-H.,  ed., Very large telecopes and their
  instrumentation Increasing the size of the isoplanatic patch size with
  multiconjugate adaptive optics.
p.~693

\bibitem[\protect\citeauthoryear{Bendek, Ammons, Belikov, Pluzhnik \&
  Guyon}{Bendek et~al.}{2012}]{bendek2012astrometry}
Bendek E.,  Ammons S.,  Belikov R.,  Pluzhnik E.,    Guyon O.,  2012, in
  Proceedings of SPIE Vol.~8442, High precision astrometry laboratory
  demonstration for exoplanet detection using a diffractive pupil telescope.
pp 844243--844243

\bibitem[\protect\citeauthoryear{Berger, Chilla, Govorkov, van Nunen \&
  Lepert}{Berger et~al.}{2012}]{berger2012towards}
Berger J.,  Chilla J.,  Govorkov S.,  van Nunen J.,    Lepert A.,  2012, in
  Proceedings of SPIE Towards a practical sodium guide star laser source:
  design for< 50 watt lgs based on opsl

\bibitem[\protect\citeauthoryear{Carrasco, Conselice C. \& Trujillo}{Carrasco
  et~al.}{2010}]{carrasco2010}
Carrasco E.,  Conselice C. J.,    Trujillo I.,  2010, MNRAS, 405, 2253

\bibitem[\protect\citeauthoryear{Carrasco, Edwards, McGregor, Winge, Young,
  Doolan, van Harmelen, Rigaut et~al.,}{Carrasco
  et~al.}{2012}]{carrasco2012results}
Carrasco E.~R.,  Edwards M.~L.,  McGregor P.~J.,  Winge C.,  Young P.~J.,
  Doolan M.~C.,  van Harmelen J.,  Rigaut F.~J.,    et~al., 2012, in
  Proceedings of SPIE Vol.~8447, Results from the commissioning of the {G}emini
  {S}outh adaptive optics imager ({GSAOI}) at {G}emini {S}outh {O}bservatory

\bibitem[\protect\citeauthoryear{Cresci, Hicks, R., Schreiber, Davies,
  Bouch\'e, Buschkamp, Genel et~al.,}{Cresci et~al.}{2009}]{cresci2009sins}
Cresci G.,  Hicks E.,  R. G.,  Schreiber N.,  Davies R.,  Bouch\'e N.,
  Buschkamp P.,  Genel S.,    et~al., 2009, The Astrophysical Journal, 697, 115

\bibitem[\protect\citeauthoryear{de Pater, Wong, Marcus, Luszcz-Cook,
  \'Adamkovics, Conrad, Asay-Davis \& Go}{de~Pater
  et~al.}{2010}]{depater2010persistent}
de Pater I.,  Wong M.,  Marcus P.,  Luszcz-Cook S.,  \'Adamkovics M.,  Conrad
  A.,  Asay-Davis X.,    Go C.,  2010, Icarus, 210, 742

\bibitem[\protect\citeauthoryear{Dicke}{Dicke}{1975}]{dicke1975phase}
Dicke R.~H.,  1975, Ap.J, 198, 605

\bibitem[\protect\citeauthoryear{d'Orgeville, Bauman, Catone, Ellerbroek, Gavel
  \& Buchroeder}{d'Orgeville et~al.}{2002}]{dorgeville2002gemini}
d'Orgeville C.,  Bauman B.~J.,  Catone J.~W.,  Ellerbroek B.~L.,  Gavel D.~T.,
    Buchroeder R.~A.,  2002, in Proceedings of SPIE Vol.~4494, Gemini north and
  south laser guide star systems requirements and preliminary designs.
pp 302--316

\bibitem[\protect\citeauthoryear{d'Orgeville, Daruich, Arriagada, Bec, Boccas,
  Bombino, Cavedoni, Collao et~al.,}{d'Orgeville
  et~al.}{2008}]{dorgeville2008gemini}
d'Orgeville C.,  Daruich F.,  Arriagada G.,  Bec M.,  Boccas M.,  Bombino S.,
  Cavedoni C.,  Collao F.,    et~al., 2008, in Proceedings of SPIE Vol.~7015,
  The {G}emini {S}outh {MCAO} laser guide star facility: getting ready for
  first light.
pp 70152P--1

\bibitem[\protect\citeauthoryear{d'Orgeville, Diggs, Fesquet, Neichel, Rambold,
  Rigaut, Serio, Araya et~al.,}{d'Orgeville
  et~al.}{2012}]{dorgeville2012gemini}
d'Orgeville C.,  Diggs S.,  Fesquet V.,  Neichel B.,  Rambold W.,  Rigaut F.,
  Serio A.,  Araya C.,    et~al., 2012, in Proceedings of SPIE Vol.~8447,
  Gemini {S}outh multi-conjugate adaptive optics ({GeMS}) laser guide star
  facility on-sky performance results

\bibitem[\protect\citeauthoryear{d'Orgeville \& McKinnie}{d'Orgeville \&
  McKinnie}{2003}]{dorgeville2003laser}
d'Orgeville C.,  McKinnie I.,  2003, in CLEO/Europe. 2003 Laser guide star
  conventional and multiconjugate adaptive optics at the {G}emini observatory:
  from one sodium laser beacon to five.
p.~737

\bibitem[\protect\citeauthoryear{Ellerbroek}{Ellerbroek}{1994}]{ellerbroek1994first}
Ellerbroek B.~L.,  1994, JOSA A, 11, 783

\bibitem[\protect\citeauthoryear{Ellerbroek \& Rigaut}{Ellerbroek \&
  Rigaut}{2001}]{ellerbroek2001methods}
Ellerbroek B.~L.,  Rigaut F.,  2001, JOSA A, 18, 2539

\bibitem[\protect\citeauthoryear{Ellerbroek, Rigaut, Bauman, Boyer, Browne,
  Buchroeder, Catone, Clark, d'Orgeville, Gavel et~al.,}{Ellerbroek
  et~al.}{2003}]{ellerbroek2003mcao}
Ellerbroek B.~L.,  Rigaut F.~J.,  Bauman B.~J.,  Boyer C.,  Browne S.~L.,
  Buchroeder R.~A.,  Catone J.~W.,  Clark P.,  d'Orgeville C.,  Gavel D.~T.,
  et~al., 2003, in Proceedings of SPIE Vol.~4839, {MCAO} for {G}emini-{S}outh.
pp 55--66

\bibitem[\protect\citeauthoryear{Elston, Raines, Hanna, Hon, Julian, Horrobin,
  Harmer \& Epps}{Elston et~al.}{2003}]{elston2003performance}
Elston R.,  Raines S.~N.,  Hanna K.~T.,  Hon D.~B.,  Julian J.,  Horrobin M.,
  Harmer C.~F.,    Epps H.~W.,  2003, in Proceeding of SPIE Vol.~4841,
  Performance of the flamingos near-ir multi-object spectrometer and imager and
  plans for flamingos-2: a fully cryogenic near-ir mos for gemini south.
pp 1611--1624

\bibitem[\protect\citeauthoryear{Esposito, Tubbs, Puglisi, Oberti, Tozzi,
  Xompero \& Zanotti}{Esposito et~al.}{2006}]{esposito2006imat}
Esposito S.,  Tubbs R.,  Puglisi A.,  Oberti S.,  Tozzi A.,  Xompero M.,
  Zanotti D.,  2006, in Proceeding of SPIE Vol.~6272, High snr measurement of
  interaction matrix on-sky and in lab..
p. 62721C

\bibitem[\protect\citeauthoryear{Feng, Taylor \& Calia}{Feng
  et~al.}{2009}]{feng200925}
Feng Y.,  Taylor L.~R.,    Calia D.~B.,  2009, Optics Express, 17, 19021

\bibitem[\protect\citeauthoryear{Fesquet, Araujo, Arriagada, Diggs, Donahue,
  d'Orgeville, Marchant, Montes et~al.,}{Fesquet
  et~al.}{2013}]{fesquet2013review}
Fesquet V.,  Araujo C.,  Arriagada G.,  Diggs S.,  Donahue J.,  d'Orgeville C.,
   Marchant C.,  Montes V.,    et~al., 2013, in AO4ELT3 Review of {G}emini
  {S}outh {L}aser {G}uide {S}tar {F}acility performance and upgrades

\bibitem[\protect\citeauthoryear{Flicker \& Rigaut}{Flicker \&
  Rigaut}{2002}]{flicker2002tilt}
Flicker R.,  Rigaut F.,  2002, in Beyond conventional adaptive optics Vol.~58,
  Tilt anisoplanatism and {PSF} retrieval in {LGS MCAO} using a predictive
  controller.
pp 377--377

\bibitem[\protect\citeauthoryear{Genzel, Eisenhauer \& Gillessen}{Genzel
  et~al.}{2010}]{genzel2010gc}
Genzel R.,  Eisenhauer F.,    Gillessen S.,  2010, Reviews of Modern Physics,
  82, 3121

\bibitem[\protect\citeauthoryear{Ghez, Salim, Weinberg, Lu, Do, Dunn, Matthews,
  Morris et~al.,}{Ghez et~al.}{2008}]{ghez2008gc}
Ghez A.~M.,  Salim S.,  Weinberg N.~N.,  Lu J.,  Do T.,  Dunn J.~K.,  Matthews
  K.,  Morris M.~R.,    et~al., 2008, The Astrophysical Journal, 689, 1044

\bibitem[\protect\citeauthoryear{Gratadour \& Rigaut}{Gratadour \&
  Rigaut}{2007}]{gratadour2007centroidgain}
Gratadour D.,  Rigaut F.,  2007, in OSA Adaptive Optics: Methods, Analysis and
  Applications Vol.~PMA4, Online centroid gain determination for lgs ao
  systems.
p.~PMA4

\bibitem[\protect\citeauthoryear{Guesalaga, Neichel, O'Neal \&
  Guzman}{Guesalaga et~al.}{2013}]{guesalaga2013vib}
Guesalaga A.,  Neichel B.,  O'Neal J.,    Guzman D.,  2013, Optics Express, 21,
  10676

\bibitem[\protect\citeauthoryear{Guesalaga, Neichel, Rigaut, Osborn \&
  Guzman}{Guesalaga et~al.}{2012}]{guesalaga2012vib}
Guesalaga A.,  Neichel B.,  Rigaut F.,  Osborn J.,    Guzman D.,  2012, Applied
  Optics, 51, 4520

\bibitem[\protect\citeauthoryear{Guyon, Bendek, Eisner, Angel, Woolf, Milster,
  Ammons, Shao et~al.,}{Guyon et~al.}{2012}]{guyon2012astrometry}
Guyon O.,  Bendek E.,  Eisner J.,  Angel R.,  Woolf N.~J.,  Milster T.~D.,
  Ammons S.~M.,  Shao M.,    et~al., 2012, The Astrophysical Journal
  Supplement, 200, 11

\bibitem[\protect\citeauthoryear{Hankla, Bartholomew, Groff, Lee, McKinnie,
  Moule, Rogers, Tiemann et~al.,}{Hankla et~al.}{2006}]{hankla2006twenty}
Hankla A.~K.,  Bartholomew J.,  Groff K.,  Lee I.,  McKinnie I.~T.,  Moule G.,
  Rogers N.,  Tiemann B.,    et~al., 2006, in Proceedings of SPIE Vol.~6272,
  20-{W} and 50-{W} solid-state sodium beacon guidestar laser systems for the
  {K}eck {I} and {G}emini {S}outh telescopes.
p. 62721G

\bibitem[\protect\citeauthoryear{Hartung, Herbst, Close, Lenzen, Brandner,
  Marco \& Lidman}{Hartung et~al.}{2004}]{hartung2004new}
Hartung M.,  Herbst T.~M.,  Close L.~M.,  Lenzen R.,  Brandner W.,  Marco O.,
   Lidman C.,  2004, Astronomy and Astrophysics, 421, L17

\bibitem[\protect\citeauthoryear{Hibon, Neichel, Prout, Rigaut, Koning, Gimeno,
  Carrasco, Winge et~al.,}{Hibon et~al.}{2013}]{hibon2013gmos}
Hibon P.,  Neichel B.,  Prout B.,  Rigaut F.,  Koning A.,  Gimeno G.,  Carrasco
  E.~R.,  Winge C.,    et~al., 2013, in AO4ELT3 First performance of the gems +
  gmos system

\bibitem[\protect\citeauthoryear{Holzl\"{o}hner, Rochester, Bonaccini~Calia,
  Budker, Higbie \& Hackenberg}{Holzl\"{o}hner
  et~al.}{2010}]{holzlohner2010sodium}
Holzl\"{o}hner R.,  Rochester S.~M.,  Bonaccini~Calia D.,  Budker D.,  Higbie
  J.~M.,    Hackenberg W.,  2010, Astronomy and Astrophysics, 510, A20

\bibitem[\protect\citeauthoryear{Hook, Jorgensen, Allington-Smith, Davies,
  Metcalfe, Murowinski \& Crampton}{Hook et~al.}{2004}]{hook2004gmos}
Hook I.,  Jorgensen I.,  Allington-Smith J.~R.,  Davies R.~L.,  Metcalfe N.,
  Murowinski R.~G.,    Crampton D.,  2004, The Publications of the Astronomical
  Society of the Pacific, 116, 425

\bibitem[\protect\citeauthoryear{Huertas-Company, Rouan, Tasca, Soucail \&
  Le~Fevre}{Huertas-Company et~al.}{2008}]{huertas2008robust}
Huertas-Company M.,  Rouan D.,  Tasca L.,  Soucail G.,    Le~Fevre O.,  2008,
  Astronomy and Astrophysics, 478, 971

\bibitem[\protect\citeauthoryear{Johnston \& Welsh}{Johnston \&
  Welsh}{1994}]{johnston1994analysis}
Johnston D.~C.,  Welsh B.~M.,  1994, JOSA A, 11, 394

\bibitem[\protect\citeauthoryear{Lu, Neichel \& Rigaut}{Lu
  et~al.}{2013}]{lu2013astrometry}
Lu J.,  Neichel B.,    Rigaut F.,  2013, in AO4ELT3 Astrometry with the
  {G}emini {M}ulti-{C}onjugate {A}daptive {O}ptics system

\bibitem[\protect\citeauthoryear{McGregor, Hart, Stevanovic, Bloxham, Jones,
  Van~Harmelen, Griesbach, Dawson et~al.,}{McGregor
  et~al.}{2004}]{mcgregor2004gemini}
McGregor P.,  Hart J.,  Stevanovic D.,  Bloxham G.,  Jones D.,  Van~Harmelen
  J.,  Griesbach J.,  Dawson M.,    et~al., 2004, in Proceeding of SPIE
  Vol.~5492, Gemini south adaptive optics imager (gsaoi).
pp 1033--1044

\bibitem[\protect\citeauthoryear{Marchetti, Brast, Delabre, Donaldson, Fedrigo,
  Frank, Hubin, Kolb et~al.,}{Marchetti et~al.}{2007}]{marchetti2007mad}
Marchetti E.,  Brast R.,  Delabre B.,  Donaldson R.,  Fedrigo E.,  Frank C.,
  Hubin N.,  Kolb J.,    et~al., 2007, in Adaptive Optics: Methods, Analysis
  and Applications Mad on-sky results in star oriented mode

\bibitem[\protect\citeauthoryear{Marchetti, Hubin, Fedrigo, Brynnel, Delabre,
  Donaldson, Franza, Conan et~al.,}{Marchetti et~al.}{2003}]{marchetti2003mad}
Marchetti E.,  Hubin N.~N.,  Fedrigo E.,  Brynnel J.,  Delabre B.,  Donaldson
  R.,  Franza F.,  Conan R.,    et~al., 2003, in Proceedings of SPIE Vol.~4839,
  Mad the eso multi-conjugate adaptive optics demonstrator.
pp 317--328

\bibitem[\protect\citeauthoryear{Moussaoui, Holzl\"{o}hner \& Hackenberg W.and
  Bonaccini~Calia}{Moussaoui et~al.}{2009}]{moussaoui2009sodium}
Moussaoui N.,  Holzl\"{o}hner R.,    Hackenberg W.and Bonaccini~Calia D.,
  2009, Astronomy and Astrophysics, 501, 793

\bibitem[\protect\citeauthoryear{Neichel, d'Orgeville, Callingham, Rigaut,
  Winge \& Trancho}{Neichel et~al.}{2013}]{neichel2013sodium}
Neichel B.,  d'Orgeville C.,  Callingham J.,  Rigaut F.,  Winge C.,    Trancho
  G.,  2013, Monthly Notices of the Royal Astronomical Society: Letters, 429,
  3522

\bibitem[\protect\citeauthoryear{Neichel, Parisot, Petit, Fusco \&
  Rigaut}{Neichel et~al.}{2012}]{neichel2012identification}
Neichel B.,  Parisot A.,  Petit C.,  Fusco T.,    Rigaut F.,  2012, in
  Proceedings of SPIE Vol.~8447, Identification and calibration of the
  interaction matrix parameters for {AO} and {MCAO} systems

\bibitem[\protect\citeauthoryear{Neichel, Rigaut, Bec, Boccas, Fesquet,
  d’Orgeville \& Trancho}{Neichel et~al.}{2011}]{neichel2011sodium}
Neichel B.,  Rigaut F.,  Bec M.,  Boccas M.,  Fesquet V.,  d’Orgeville C.,
  Trancho G.,  2011, in AO4ELT2 Sodium {P}hoton {R}eturn, spot elongation and
  fratricide effect: {F}irst on-sky results with {GeMS}.
p.~54

\bibitem[\protect\citeauthoryear{Neichel, Rigaut, Serio, Arriagada, Boccas,
  d'Orgeville, Fesquet, Trujillo, Rambold, Galvez et~al.,}{Neichel
  et~al.}{2012}]{neichel2012science}
Neichel B.,  Rigaut F.,  Serio A.,  Arriagada G.,  Boccas M.,  d'Orgeville C.,
  Fesquet V.,  Trujillo C.,  Rambold W.~N.,  Galvez R.~L.,    et~al., 2012, in
  Proceedings of SPIE Vol.~8447, Science readiness of the gemini {MCAO} system:
  {GeMS}

\bibitem[\protect\citeauthoryear{Ragazzoni, Diolaiti, Farinato, Fedrigo,
  Marchetti, Tordi \& Kirkman}{Ragazzoni et~al.}{2002}]{ragazzoni2002multiple}
Ragazzoni R.,  Diolaiti A.,  Farinato J.,  Fedrigo E.,  Marchetti E.,  Tordi
  M.,    Kirkman D.,  2002, Astron. Astrophys, 396, 731

\bibitem[\protect\citeauthoryear{Rigaut}{Rigaut}{2000}]{rigaut2000ttta}
Rigaut F., , 2000, Compensation of the null modes with MCAO,
  \url{http://www.gemini.edu/sciops/instruments/mcao/pdf/PlateScaleModes.pdf}

\bibitem[\protect\citeauthoryear{Rigaut}{Rigaut}{2001}]{rigaut2001glao}
Rigaut F.,  2001, in Beyond conventional adaptive optics Vol.~58, Ground
  conjugate wide field adaptive optics for the elts.
p.~11

\bibitem[\protect\citeauthoryear{Rigaut, Neichel, Boccas, d'Orgeville,
  Arriagada, Fesquet, Diggs, Marchant et~al.,}{Rigaut
  et~al.}{2012}]{rigaut2012gems}
Rigaut F.,  Neichel B.,  Boccas M.,  d'Orgeville C.,  Arriagada G.,  Fesquet
  V.,  Diggs S.~J.,  Marchant C.,    et~al., 2012, in Proceedings of SPIE
  Vol.~8447, Gems: First on-sky results

\bibitem[\protect\citeauthoryear{Rigaut, Neichel, Boccas, d'Orgeville, Vidal,
  van Dam, Arriagada, Fesquet et~al.,}{Rigaut et~al.}{2014}]{rigaut2013review}
Rigaut F.,  Neichel B.,  Boccas M.,  d'Orgeville C.,  Vidal F.,  van Dam M.~A.,
   Arriagada G.,  Fesquet V.,    et~al., 2014, Monthly Notices of the Royal
  Astronomical Society, 437, 2361

\bibitem[\protect\citeauthoryear{Rigaut, Neichel, Dorgeville, Gratadour,
  Boccas, Trancho, Bec, Trujillo, Edwards \& Carrasco}{Rigaut
  et~al.}{2011}]{rigaut2011gems}
Rigaut F.,  Neichel B.,  Dorgeville C.,  Gratadour D.,  Boccas M.,  Trancho G.,
   Bec M.,  Trujillo C.,  Edwards M.,    Carrasco R.,  2011, in AO4ELT2 Gems
  sees star light

\bibitem[\protect\citeauthoryear{Rigaut \& Roy}{Rigaut \&
  Roy}{2001}]{rigaut2000science}
Rigaut F.,  Roy J.-R., , 2001, The science case for the multi-conjugate
  adaptive optics system on the Gemini South telescope,
  \url{http://maumae.net/gems/MCAO-SC-V2.0.2.pdf}

\bibitem[\protect\citeauthoryear{Rigaut, Ellerbroek \& Flicker}{Rigaut
  et~al.}{2000}]{rigaut2000principles}
Rigaut F.~J.,  Ellerbroek B.~L.,    Flicker R.,  2000, in Proceedings of SPIE
  Vol.~4007, Principles, limitations, and performance of multiconjugate
  adaptive optics.
pp 1022--31

\bibitem[\protect\citeauthoryear{Rochester, Otarola, Boyer, Budker, Ellerbroek,
  Holzl\"{o}hner \& Wang}{Rochester et~al.}{2012}]{rochester2012sodium}
Rochester S.~M.,  Otarola A.,  Boyer C.,  Budker D.,  Ellerbroek B.,
  Holzl\"{o}hner R.,    Wang L.,  2012, JOSA-B, 29, 2976

\bibitem[\protect\citeauthoryear{Rodriguez, Neichel, Hartung, Haywards,
  Christou, Rigaut, Guzman \& Guesalaga}{Rodriguez
  et~al.}{2011}]{rodriguez2011vibration}
Rodriguez I.,  Neichel B.,  Hartung M.,  Haywards T.,  Christou J.,  Rigaut F.,
   Guzman D.,    Guesalaga A.,  2011, in AO4ELT2 Vibration characterization and
  mitigation at the {G}emini-{S}outh telescope

\bibitem[\protect\citeauthoryear{Trancho, Bec, Artigau, d'Orgeville, Gratadour,
  Rigaut \& Walls}{Trancho et~al.}{2008}]{trancho2008gemini}
Trancho G.,  Bec M.,  Artigau E.,  d'Orgeville C.,  Gratadour D.,  Rigaut
  F.~J.,    Walls B.,  2008, in Proceedings of SPIE Vol.~7016, The
  {G}emini-{S}outh {MCAO} operational model: insights on a new era of telescope
  operation

\bibitem[\protect\citeauthoryear{van Dam}{van
  Dam}{2005}]{vandam2005centroidgain}
van Dam M.,  2005, JOSA-A, 22, 1509

\bibitem[\protect\citeauthoryear{van Dam, Bouchez, Conan \& McLeod}{van Dam
  et~al.}{2013}]{vandam2013glao}
van Dam M.,  Bouchez H.,  Conan R.,    McLeod A.,  2013, in AO4ELT3 Wavefront
  reconstruction for a natural guide star ground layer adaptive optics system
  on the giant magellan telescope

\bibitem[\protect\citeauthoryear{V\'{e}ran \& Herriot}{V\'{e}ran \&
  Herriot}{2000}]{veran2000centroidgain}
V\'{e}ran J.,  Herriot G.,  2000, JOSA-A, 17, 1430

\bibitem[\protect\citeauthoryear{Vidal, Neichel, Rigaut, Carrasco, Winge,
  Pessev, Serio, Arriagada et~al.,}{Vidal et~al.}{2013}]{vidal2013gems}
Vidal F.,  Neichel B.,  Rigaut F.,  Carrasco R.,  Winge C.,  Pessev P.,  Serio
  A.,  Arriagada G.,    et~al., 2013, in AO4ELT3 Gems: from the on-sky
  experimental system to science operation. the {AO} point of view

\bibitem[\protect\citeauthoryear{Wizinowich}{Wizinowich}{2012}]{wizinowich2012progress}
Wizinowich P.,  2012, in Proceedings of SPIE Vol.~8447, Progress in laser guide
  star adaptive optics and lessons learned

\bibitem[\protect\citeauthoryear{Wright, Larkin, Law, Steidel, Shapley \&
  Erb}{Wright et~al.}{2009}]{wright2009dynamics}
Wright S.,  Larkin J.,  Law D.,  Steidel C.,  Shapley A.,    Erb D.,  2009, The
  Astrophysical Journal, 699, 421

\bibitem[\protect\citeauthoryear{Young, McGregor, van Harmelen,  \&
  Neichel}{Young et~al.}{2012}]{young2012odgw}
Young P.~J.,  McGregor P.~J.,  van Harmelen J.,     Neichel B.,  2012, in
  Proceedings of SPIE Vol.~8447, Using odgws with gsaoi: Software and firmware
  implementation challenges

\end{thebibliography}

\label{lastpage2}

\end{document}